# Is there a "Moore's law" for quantum computing?

Olivier Ezratty [1]

[1] author of the Understanding quantum technologies book and cofounder of the Quantum Energy Initiative, Paris, France,

olivier@oezratty.net @olivez

There is a common wisdom according to which many technologies can progress according to some exponential law like the empirical Moore's law that was validated for over half a century with the growth of transistors number in chipsets. As a still in the making technology with a lot of potential promises, quantum computing is supposed to follow the pack and grow inexorably to maturity. The Holy Grail in that domain is a large quantum computer with thousands of errors corrected logical qubits made themselves of thousands, if not more, of physical qubits. These would enable molecular simulations as well as factoring 2048 RSA bit keys among other use cases taken from the intractable classical computing problems book. How far are we from this? Less than 15 years according to many predictions. We will see in this paper that Moore's empirical law cannot easily be translated to an equivalent in quantum computing. Qubits have various figures of merit that won't progress magically thanks to some new manufacturing technique capacity. However, some equivalents of Moore's law may be at play inside and outside the quantum realm like with quantum computers enabling technologies, cryogeny and control electronics. Algorithms, software tools and engineering also play a key role as enablers of quantum computing progress. While much of quantum computing future outcomes depends on qubit fidelities, it is progressing rather slowly, particularly at scale. We will finally see that other figures of merit will come into play and potentially change the landscape like the quality of computed results and the energetics of quantum computing. Although scientific and technological in nature, this inventory has broad business implications, on investment, education and cybersecurity related decision-making processes.

## INTRODUCTION

Gordon Moore's law successfully described the growth rate of transistor per chipset between 1965 and now. Many signs are showing it is reaching some limits, but more on other dimensions like with transistors density, computing power per area, computing cores power, number of cores and clock speed.

For about a decade now, quantum computing scientists and technologists have tried to identify various exponential progress laws similar to Moore's law, particularly on qubit numbers, qubit fidelities and other figures of merit. One can indeed wonder whether history will repeat itself in the quantum computing world. Understanding how such laws could work and making forecasts is not just about gut feelings and a naive optimistic view on technology progress determinism. It requires a mix of understanding of the scientific and technology challenges faced by quantum physicists and quantum computers creators but also of the underlying economics of this emerging business. Predicting some Moore's law for quantum computing progress is a highly cross-disciplinary challenge.

Why may it be relevant to find some empirical laws on the development of quantum computing? There are at least a couple business and scientific reasons.

One is to assess when quantum computing will become a real business outside the proof of concept zone it is into nowadays. It is useful for investors, governments, and customers to have a better clue on their own quantum computing agenda, despite all the related uncertainties.

For example, it may influence the decision making process on launching quantum computing related educational programs and on the way to balance scientific fundamental research and technology development investments.

Another reason is linked to cybersecurity. It is common practice to exaggerate the quantum computing threat on current cybersecurity based on RSA public key infrastructures. Surveys are regularly done polling 40 worldwide experts in quantum physics and quantum information science[1].

The expectation for seeing a quantum computing breaking RSA 2048 keys is averagely "in 15 years" with a Gaussian curve of response around this timeframe showing a broad discrepancy of opinions among experts. Some would like to predict the future of quantum computing based only on the billions of dollars invested by governments around the world (at least $15B so far) or in venture capital (about $7B so far), on top of the large information technology companies' investments (IBM, Amazon, Google, Microsoft, Intel, Alibaba, …) which is probably a bit shortsighted[2]. Making such predictions requires a strong understanding of the science and technology behind quantum computing and classical computing, and from hardware to software. Quantum computing is a long term quest still requiring a lot of fundamental research work, including for the many startups operating in that field. The aim of this paper is to unfold this challenge piece by piece. And the response to the question asked in the paper title will be quantum in nature: yes and no!

## MOORE'S LAW IN CLASSICAL COMPUTING

As a disclaimer, many scientists consider Moore's law as a questionable misnomer. It is not a law per se. It is not related to some immutable underlying physical laws like those governing classical or quantum physics. But it's called a law, that's the way it is, although it is highly empirical.

Moore's law was a sort of exponential regression used to predict the rate of growth of the number of transistors in a chipset, doubling every 24 or 18 months [3]. It was based on a sampling made with only five data points ranging from 1960 to 1965, in the very early years of the history of integrated circuits production. It had been invented by Jack Kilby from Texas Instrument in 1958 and first produced in 1960[4]! It bet on a growth of transistor density more than on an increase in chipset size, in days when a regular wafer was only one inch large. Nowadays, wafers are 12 inches large (30 cm) and can accommodate hundreds if not thousands of chipsets depending on their size, or just one large chipset, like Cerebras' giant CS-2 wafer-scale chipset manufactured by TSMC.

Moore's law was later observed with many other technology (more or less) exponential growth patterns such as with digital storage data capacity, speed and latency[5], supercomputers computing power, wired and wireless networking and telecommunications, DNA sequencing cost per human genome, solar panels yield or prices and the likes. It can even be attributed to some progress in algorithms design, in the high-performance computing world[6].

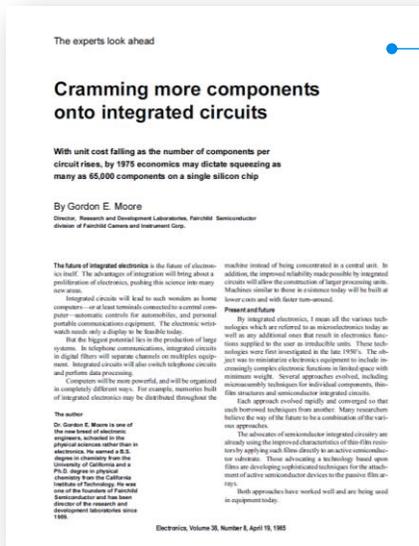
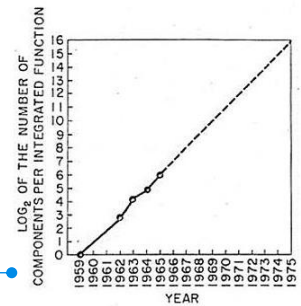
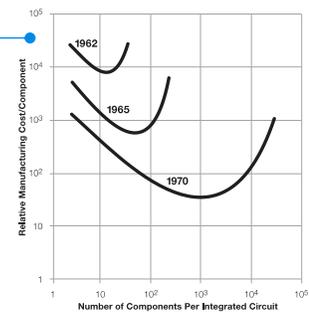

empirical observation from 1965
the complexity of integrated circuits doubles every 18 months

"complexity"
= # of transistors on a chip

also, an economics driven law

many derivatives with:
- transistors density
- cost / transistor
- supercomputing power
- storage capacity
- cost of storage / GB
- networking speed
- CMOS imaging sensors resolution
- human genome sequencing cost

Figure 1 : Gordon Moore's law was a technology-based and also an economical-based prediction. Source: Gordon Moore paper[3] and Olivier Ezratty.

## Moore's law economic rationale

Moore's law was aligned with an economic rationale, related to the cost optimization of transistors manufacturing[7]. It was not obvious back then. One key concern was how the semiconductor industry could improve manufacturing yield problem although in 1965, the industry was already optimistic about it.

In real life, as Intel launched new improved density nodes, yield was falling and improving later. The economical acceleration drive was the massification of chipsets usages, driven initially by the micro-computing revolution starting in the late 1970s.

The timing of Gordon Moore's paper was important. It was written when he worked at Fairchild Semiconductors, only 5 years after the production of the integrated circuit and 6 years before Intel created its first microprocessor, the 4004. It was an era of relatively fast technological progress. The Moore's law nickname was created after Moore's paper was published, by Carver Mead, a Professor at Caltech and friend of Gordon Moore, who passed away on March 24th, 2023 while this paper was being finalized.

Moore's paper, shown in Figure 1, was particularly visionary, including this statement: "*integrated circuits will lead to such wonders as home computers or at least terminals connected to a central computer, automatic controls for automobiles, and personal portable communications equipment*". Fifteen years before the other famous Bill Gates willingness to put a "*personal computer on every desk and in every home*" (… running Microsoft software).

## Dennard, Koomey and dark silicon

Moore's law is said to have ended years ago. It's not entirely exact if we stick to its original definition as shown in Figure 4. The number of transistors per chipset continued to grow steadily past 2020. Four decades after Moore's paper, 2006 signaled the end of another empirical law, the so-called **Dennard scale** according to which there was a stability of power consumption per chipset surface as density improved, voltage was reduced and clock speed increased (see Figure 3)[8]. It was derived as **Koomey's law**, created in 2010, according to which the energy efficiency of computers doubled about every 18 months, faster than the initial Moore's law 24 months period, but only every 2 years starting in 2000, in relation with the end of Dennard's scale[9].

Starting at 65 nm nodes, thermal dissipation due to transistors leakage increased within chipsets, generating the so-called "dark silicon" phenomenon which prevents a chipset from running all its components simultaneously at full clock speed. Also, a transistor voltage threshold around 0.4V prevents a power consumption decrease per transistor. Therefore, chipset clock speeds have nearly stalled during the last 15 years. It is easy to understand why, as a chipset power consumption increases as the cube of its clock frequency above 1 GHz[10]. In the end, the real price of useful transistors is not decreasing that fast and may even be increasing as density continues to grow, and also due to the growing cost of designing more complicated chipsets.

This didn't prevent the semiconductor industry from continuing to invest down the road of Moore's law with regards to the number of transistors in a chipset and their density. The recent records are of about 50 billion transistors for the latest generation Nvidia H100 GPGPU (which packs two such chipsets next to some High Bandwidth Memory (HBM3), all in the same package)[11] and the Cerebras CS-2 21.5x21.5cm huge wafer-scale processor with its 2.6 trillion transistors, 850.000 cores and ultra-fast 40 GB on-chip memory[12].

TSMC is now manufacturing "3 nm" chipsets which, according to the company, offer "*up to 70% logic density gain, up to 15% speed improvement at the same power and up to 30% power reduction at the same speed as compared with N5 technology*"[13]. This is the "More Moore" roadmap.

However, the hard truth about chipsets integration is that "3 nm" nodes processors do not have any transistor features of this size[14]. This dimension was supposed to be the gate pitch. Transistors are much larger, with sizes ranging between 20 and 40 nm, even in the future sub 1 nm node ranges. 10 nm, 7 nm, 5 nm and 3 nm node range names are just labelling from the semiconductor industry to describe consecutive generation transistor densities[15]. Since 2017, the related transistor feature sizes have no connection with this labelling, including gate pitch, as shown in Figure 2[16]. Only the gate oxide can have a thickness under one nanometer, and it's been the case since the late 2000s.

Figure 2: the real transistor feature sizes per generation showing that 3 nm, 2 nm and below do not correspond to any real size in transistor designs in horizontal features. Transistor size is not significantly changing from one generation to the other, validating the "end of Moore's law" claim. The right way to describe these nodes would be a number scheme like G48M24T1 with a gate pitch of 40 nm, a metal pitch of 24 nm and one layer of transistors for 3 nm nodes. The 0.5 nm in the table above would become G38M16T6. It is of course more complicated that 0.5 nm! Source: IEEE IRDS 2022 More than Moore report[17].

Starting in 2021, having reached the limit of horizontal integration at the 40 nm / 16 nm (gate pitch / metal pitch) level, CMOS logic manufacturers will begin to stack several layers of transistors on top of each other with up to 6 layers by 2037 with the 3DVLSI technology.

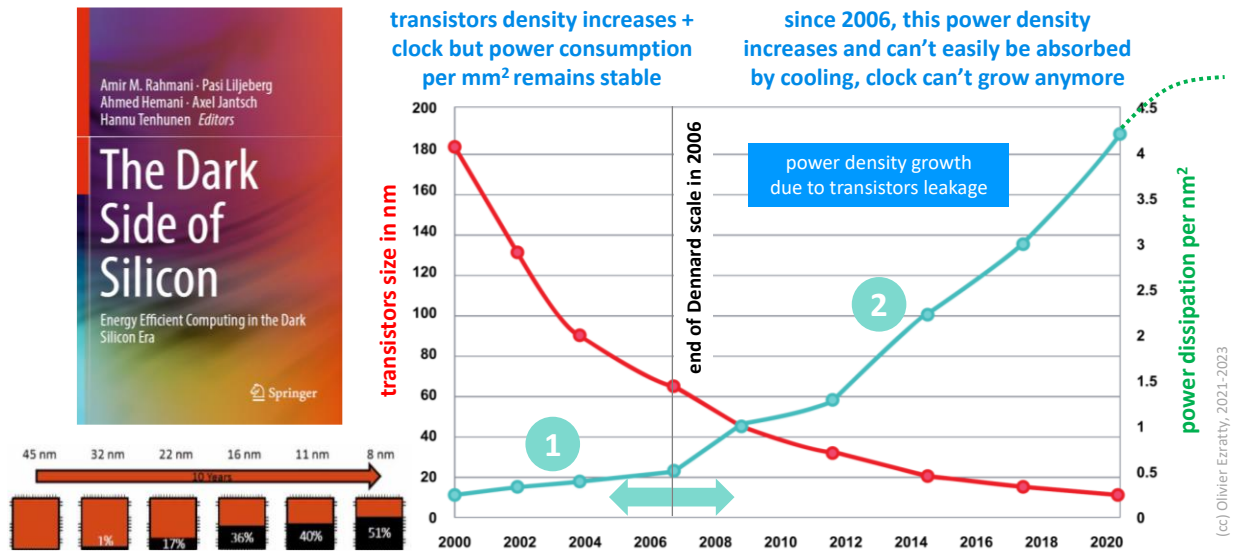

Figure 3: Dennard's scale and the dark side of silicon. Various sources like Hennessy[18].

Otherwise, many solutions have been invented to circumvent the heating problem like using multicore architectures (IBM, Intel [19] [20]), slowing down the speed of processors like with the ≈1 GHz base clock of Nvidia GPGPUs, deactivating chipset components selectively, mixing high performance / high power and low performance / low power cores like with smartphone arm-based chipsets and recent Intel PC chipsets, DSA (domain specific architectures) with efficient and dedicated functions engraved in silicon like NPUs (neural processing units) implementing matrix multiplications used for neural networks, video and audio codecs, digital signals processing, networking, security and the likes.

Some PC chipsets can have clocks reaching 6 GHz, but it is usually limited to a couple operating cores. It is the case of the Intel 13900K introduced in January 2023 with its 8 fast cores and 16 slow cores where only two fast cores can run simultaneously at the 6 GHz turbo-speed[21] ! This CPU can also run at a record overclocked speed of 9 GHz thanks to some rather impractical liquid helium cooling[22].

**Multicore chipsets**

The number of cores of typical CPUs has not grown significantly in the last half-dozen years, one of the reasons being the **Pollack rule**, devised by an Intel engineer in the late 1990's, according to which the growth of performance linked to microarchitecture advances like multicore CPUs is proportional to the square root of the chipset area while the power consumption increases linearly with the same area. NPUs (specialized neural processing units like Google TPU, Graphcore BOW IPU processors, Cerebras CS-2) and GPGPU (Nvidia V100/A100/H100) still have a very large number of cores working in parallel, but these use a different architecture than with typical CPUs, with dedicated matrices/vectors multipliers/adders driven by an external logic and implementing in "core clusters" the same sets of instructions.

Their speed is highly dependent on memory access, thus the growth of internal SRAM (which is very costly) or surrounding DRAM using data-exchanges with HBM3E protocols, providing a high-memory bandwidth reaching a whopping 3.35 TB/s per GPGPU [23]. HBM4E bandwidth is planned to reach 10 TB/s by 2030[24]. This is much faster than any data transfer from a classical computer to a quantum computer, even in the long term.

Gordon Moore's law was mainly enabled by manufacturing process improvements, particularly at the lithography step, using multi-patterning and then higher ultra-violet frequencies to increase resolution. It led to the leadership of ASML (The Netherlands) who is now a monopoly with its EUV (extreme-ultraviolet) photolithography machines [25].

**More than Moore**

Since about 2015, the semiconductor industry is positioning the current era in a new **More than Moore** period which privileges chipsets hybridization over density improvements[26]. This is a concept that is currently implemented in smartphone and other computing architectures with "SIP" (systems in package), multi-chip modules (using a silicon interposer with several "chiplets" on a passive silicon carrier placed on an organic substrate[27]) assembling various chipsets designed with different manufacturing processes, 2D, 2.5D and 3D, and integrated in highly compact packages. You can now embed into a SIP a classical CMOS chipset, various sensors, analog/digital RF components and even optical components[28].

The potential end of Moore's law effects is frequently used as a motivation to develop quantum computers. I'm not sure it comes from the classical semiconductor and computers industry. Rather, it is an argument coming from the quantum computing ecosystem.

Other options are also investigated in many other directions, belonging to the broad field of "unconventional computing", with superconducting logic providing much higher speeds and lower power consumption, digital annealing in CMOS, Fast-Fourier Transform implemented in silicon [29] [30], reversible computing with a much lower power footprint, neuromorphic computing also with a lower power footprint, spintronics and memristors used in the former, optical computing and other avenues[31]. Each of these technologies are still immature. For example, superconducting logic can't handle memory and must be complemented by another relatively immature technology, spintronic. Reversible computing could theoretically consume much less energy, but at the expense of much larger circuits and slower speed.

Some of these technology paths may become enablers of some exponential progress in quantum computing as we'll see later, particularly in the control electronics domain.

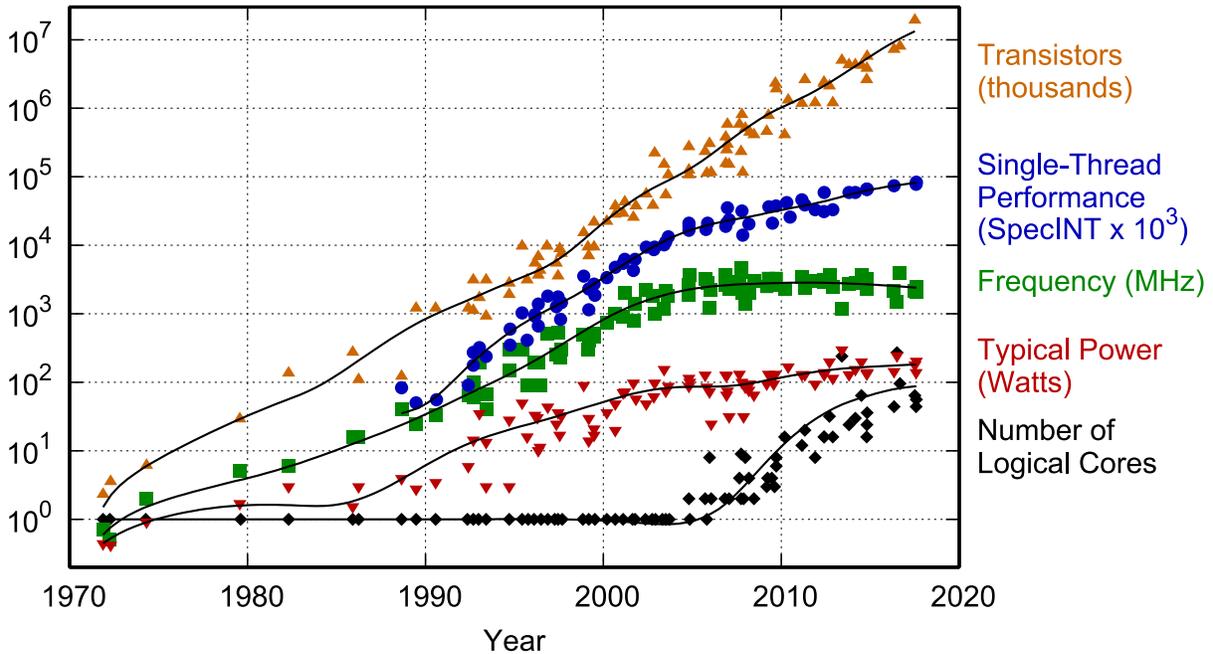

Figure 4: how microprocessor figures of merits progress slowed down with single thread performance, clock speed, and number of logical cores, in relation with total power consumption. Still, Moore's law related to the number of transistors per chipset is always valid. Source: Karl Rupp[32].

**QUANTUM COMPUTING FIGURES OF MERIT**

Now, onto quantum computers and trying to respond to the post title with identifying whether some trends observed in quantum computing performance and engineering are similar to the trend known as "Moore's law" and could be projected in the future.

One first challenge is to understand the origins of the power of quantum computing and its potential speedups. It is often attributed to some form of massive parallelism but it's too classical an image to describe what is happening within a quantum computer which is highly analog in nature. Then, a quantum computer speedup is highly variable, dependent on the algorithm type, on where the data comes from, and how results are generated and retrieved from the quantum computer. Many quantum algorithms show a modest polynomial speedup while a few bring an exponential one. There is also some significant overhead to account for, with quantum error mitigation techniques used in noisy quantum computers ("NISQ" for noisy intermediate scale quantum)[109] and quantum error correction codes in future fault-tolerant quantum computers ("FTQC")[33]. Also, most quantum algorithms are hybrid and have a classical part.

Our second challenge is that we're not dealing with a single technology like the integrated circuits of the 1960's and 1970's dominated by CMOS silicon-based transistors. There is a large variety of qubit technologies which are miles apart in physical aspects, operational and miniaturization challenges. Neutral atoms in vacuum are entirely different from superconducting and electron spin qubits, and photon-based qubits are also weird beasts, *aka* flying qubits, compared to the other types of qubits which are static in location. Contrarily to solid state qubits, photons don't have much of a decoherence problem but are harder to generate, control and detect in a deterministic way. Even though it is still a fundamental research topic, flying electrons are also investigated as flying qubits or interconnect qubits to create connection between other spin and static qubits.

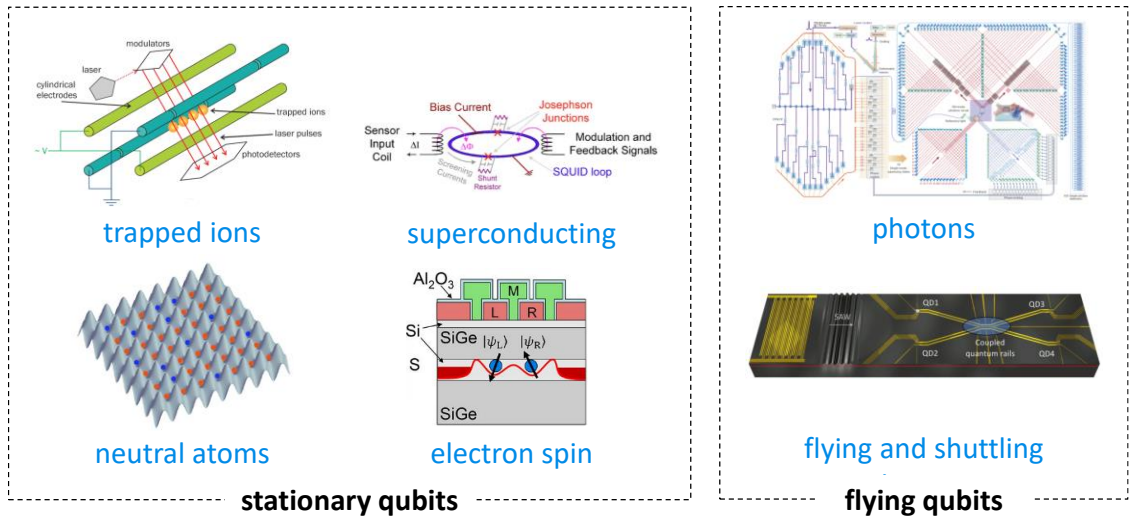

Figure 5: the main types of qubits given each class has it own many variants. Compilation: Olivier Ezratty.

Then, assembling a useful quantum computer and scaling it is a cross-disciplinary challenge involving many domains: qubit chipsets manufacturing for all technologies aside from neutral atom qubits and a lot of various enabling technologies (analog and digital electronics, data-transit and computing, software, error correction codes, cryogeny, lasers, cabling, etc.).

Economics are also different since the timescale of "commoditization" of quantum computers is quite far contrarily to the microcomputing revolution of the mid-1970s which started less a decade after Moore coined his empirical prediction. And it is the case, even when accounting for the wildest market predictions from McKinsey[34] and BCG[35] who tend to describe customer business value than computing industry revenues, creating significant confusion [36].

The microcomputing revolution was one of the economic drivers of Moore's law between the 1980s until about 2010, when the household PC equipment rate grew from a couple percent to above 85% in most developed countries (2.8% in 1980 to 93.3% in 2023 in the USA)[37]. Then, the smartphone era expanded the reach of computing to an even greater share of the worldwide population. Far from these two macro-technology trends, the economics of quantum computing will probably be closer to the very specialized high-performance computing market.

Since the advent of the first small-scale quantum computers in the early 2000s, researchers and analysts are plotting charts with number of qubits and other qubit figures of merit, trying to showcase a similar law with quantum computers. Qubit numbers replaced transistor count. Qubit density is not yet at play and is not really improving. The real priority is to make better qubits even though qubit counts are still increasing in the commercial realm roadmaps without taking much care about qubit fidelities[38].

## Qubit count

In 2003, Geordie Rose, then CTO and cofounder of D-Wave, proposed his own prediction on the rate of progress of quantum computers, based only on the number of physical qubits[39]. It fared relatively well for D-Wave quantum annealers. Their first prototype, Calypso, had 4 qubits in 2007, their first commercial product had 128 qubits in 2012 and in 2020, D-Wave Advantage generation reached 5000 qubits.

Until 2020, it looks like a good exponential curve as shown in Figure 6. Since then, it has nearly stalled. D-Wave is planning a successor to the D-Wave Advantage named Advantage 2, with about 7000 qubits and a better connectivity between them, from 1 to 15, to 1 to 20 in what they called logical qubits which are close assemblies of physical qubits. It makes it easier to handle the "embedding" of a problem into their system, meaning, mapping an optimization problem onto the qubit layout. In June 2022, D-Wave released a prototype chipset for this generation with 576 qubits using their new Zephyr topology[40].

IBM's progress with its number of qubits follows a similar path starting in 2015 with 1 qubit and now with 433 qubits, heading to 1121 this year (2023) and later, above 4000 physical qubits. The exponential is perfect. We'll see of course that it is not enough to create useful systems.

Besides these two vendors, qubit numbers progress has been quite sluggish. Google announced 72 qubits in 2018 with Bristlecone, didn't really test/use it[41], then used its 53 qubits Sycamore processor in 2019 for its famous supremacy experiment[42], and is now back at... 72 qubits since July 2022 with a Sycamore update. Rigetti announced 128 qubits in 2018[43] and is now at 80 with its Aspen-M2 and M3 generation[44]. IonQ has been stalled at around 20-32 qubits for several years. And on and on. Figure 6 shows the rate of progress with D-Wave, IBM, Google and Rigetti.

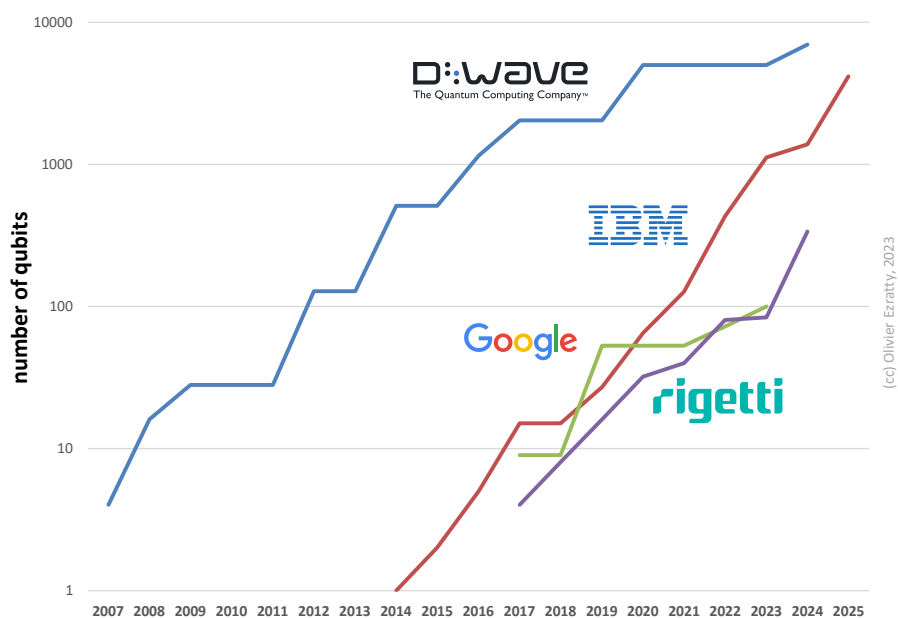

Figure 6: evolution of the number of physical qubits with D-Wave, IBM, Google and Rigetti. Compilation: Olivier Ezratty.

In 2019, Hartmut Neven from Google suggested another empirical law, bearing his name, according to which quantum computers computing power would grow by a double exponential law. The first exponent comes from the exponential computing space advantage coming from N qubits as N grows, and the second, from the exponential growth of N over time[45]. Looking at Figure 6 seems to validate this trend. This double exponential is misleading since the space advantage of N qubit doesn't automatically turn into an exponential speed advantage with quantum algorithms, even with perfect qubits. Only a few quantum algorithms enable an exponential speedup, and they require a significant space and time overhead with quantum error correction.

A qubit number is also not the single indicator of computing power. Even in the NISQ era of quantum computers with over 50 noisy qubits, you still need high quality qubits with long coherence times and low error rates to do something useful and envision reaching a speed-based quantum advantage, or just, to make some computing generating good results[46]. The usual figures of merit are the $T_1$ and $T_2$ corresponding to qubits amplitude and phase stabilities, and two key numbers: the two-qubit error rates and the qubit readout error rates. Another less used but as important figure of merit is the qubit initialization error rate. If you can't reset perfectly your qubit in the $|0\rangle$ state from the start, things won't go well afterwards during your computation. State preparation and measurement (SPAM) metrics are also used, mostly by trapped ions vendors. It complements qubit gate fidelities that sit in between preparation and measurement.

*Findings*: as we'll see later, qubit count grows faster than qubit fidelities, making most large qubit counts unusable at the exception of analog quantum computers.

*Pending question*: will research and vendors catch up with qubit fidelities and create architectures that have stable fidelities as the qubit count grows?

**Qubit lifetimes**

In the scientific literature, you can see progress on these various qubit figures of merit being plotted until 2016 and sometimes up to 2020. Let's look at these various numbers, first with superconducting qubits coherence time ($T_1$). You can observe the scale of process in two decades, with $T_1$ growing from a couple nanoseconds to over a millisecond, so above five orders of magnitude.

I added to the chart in Figure 7 which is from 2020 a couple additional points with Google Sycamore's 2019 $T_1$, University of Maryland's fluxonium record $T_2$ of 1.48 ms in 2021[47], and IBM's 127 qubit Sherbrooke processor with its excellent 300 µs $T_1$ [48]. But it is quite hard to tell if these $T_1$ and $T_2$ are computed in a consistent way. For example, IBM's Sherbrooke transmon $T_1$ is achieved with a 127 qubit chipset while most other $T_1$ were laboratory implementations with only a few qubits.

Other types of qubits show sometimes amazing coherence times, like SiC centers (silicon carbide defects) which showcase $T_1$ in seconds[49]. But these are still isolated qubits with no single or two-qubit gate fidelity numbers.

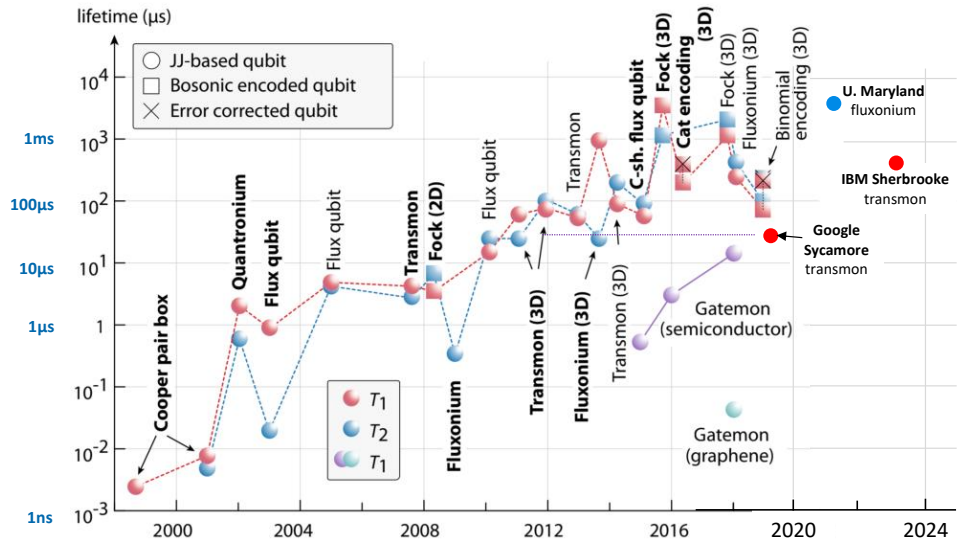

Figure 7: evolution of superconducting lifetime over time. Source: Morten Kjaergaard et al[50].

*Findings:* with superconducting qubits, coherence times have made amazing progress with six orders of magnitude improvement ranging from the nanosecond to the millisecond. It has stabilized in a couple years, particularly with commercial products but, meanwhile, gate times and fidelities have also improved significantly, extending these systems computing capabilities. In that realm, fluxonium superconducting qubits look promising.

*Pending question*: will researchers and vendor continue to improve these coherence times? What are the equivalent figures of merits for other types of qubits than superconducting qubits (photons, neutral atoms, silicon, NV centers, topological, …)?

**Qubit fidelities**

Now, let's look at qubit fidelities ranging from qubit initialization, qubit single and two-qubit gates and qubit readout fidelities. Some charts exist but there is no single way of benchmarking these fidelities although randomized benchmarking seems to be the dominant technique used there. I didn't find many similar charts with all types of qubits. The one in Figure 8 shows progress with ion trap and superconducting qubits two-qubit gate fidelities using vendor data. In some cases, like with near-field microwave control of ion traps, there are two few points to drive any relevant observation [51].

Figure 9 plots some two-qubit gate error rates of industry vendor quantum computers against their number of qubits, based on vendors data. It shows interesting trends. These qubit fidelities are centered around 1% and it seems to slightly decrease for IBM as they grow the number of qubits of their processors.

For IBM, in green, the arrow of time corresponds more or less to a bottom to top direction, corresponding to the number of qubits.

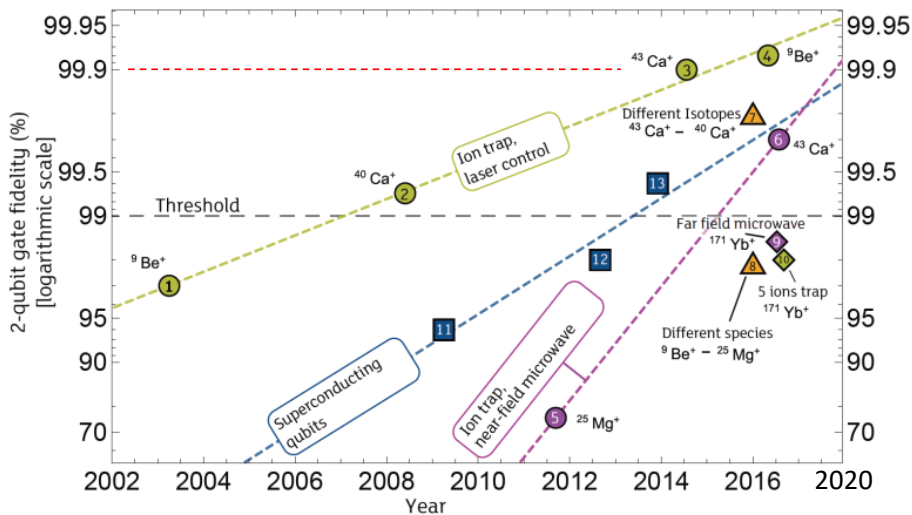

Figure 8: two-qubit error rates with ion traps and superconducting qubits. Source: Amir Fruchtman and Iris Choi, University of Oxford[52].

For a given generation (27 qubits, 65 qubits, 127 qubits), some improvements can be identified from right to left. IBM's 433 qubit Osprey that was revealed in November 2022 is not in this chart since IBM has not yet released any related qubit fidelity data. It will probably follow the direction of the green arrow in Figure 9.

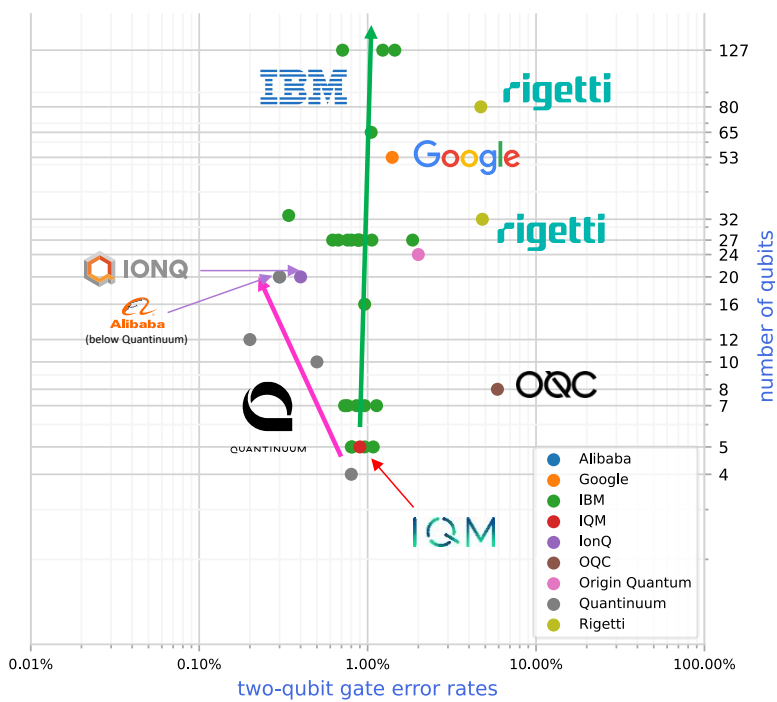

Figure 9: scatter plot of qubit numbers and two-qubit gate error rates for commercial vendors. Source: vendor fidelities numbers compiled by Olivier Ezratty as of March 2023.

Other superconducting qubit vendors showcase much lower fidelities (Rigetti, OQC) or average fidelities (IQM, Google). Ion trap processors have better and growing fidelities although with a lower number of qubits. The scalability of this type of qubit remains an open question given its slow progress over the last few years.

As part of its METRIQ project, the Unitary Fund consolidates QPU performance data from published papers[53]. Its two-qubit fidelities chart in Figure 10 shows the evolution of the data over time, not number of qubits. As usual, best-in-class fidelities come from trapped ions qubits.

The most recent IBM data in the chart corresponds to 5-qubit QPUs, with fidelities around 1%. These are somewhat consistent with the data consolidated from vendor numbers in Figure 9. In both cases, these benchmarks are implemented using the randomized benchmarking method.

Two key technological advances are in the waiting: one is to continue to improve qubit quality, the other is to increase their number while keeping this improved quality. That's an enormous challenge. IBM's Heron 133 qubit processor is supposed to be released in 2023 and to showcase 99.9% two-qubit gate fidelities. Superconducting fluxonium qubits look also promising with best-in-class vendor results coming from Alibaba with 99.7% two-qubit gate fidelities, although with only 20 qubits so far, with some scaling challenges related to how qubits are interconnected[54] [55] [56].

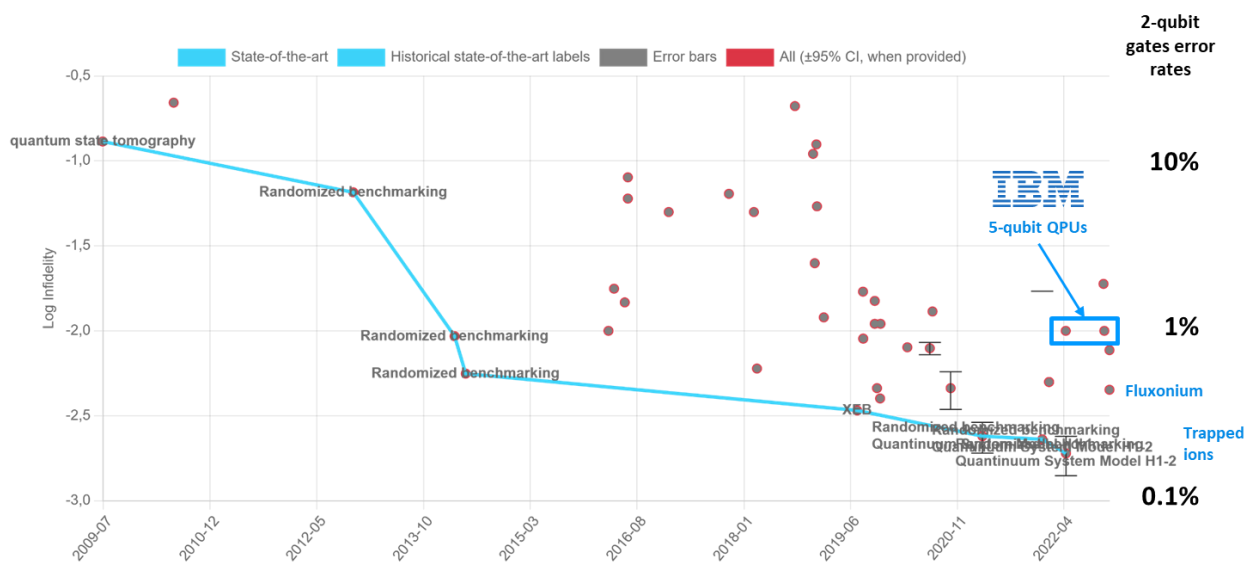

Figure 10: chart of two-qubit gates fidelities over time as consolidated by the Unitary Fund in its Metriq initiative[57]. A couple additions by Olivier Ezratty (in blue).Now, let's zoom out Figure 9 and include what would be needed to implement fault-tolerant quantum computing, as shown in Figure 11.

Besides IBM, the only other vendors making such promises are designing bosonic-code qubits, like Alice&Bob and Amazon with their cat-qubits. These qubits are self-correcting flip errors. The remaining phase errors must be corrected with quantum error correction codes, landing these qubits directly in the fault-tolerant quantum computing generation. As a result, much fewer physical qubits would be required to create typical logical qubits in the FTQC realm[58]. This promising technological development is still in its early stage. Another proposal by Fujitsu, Osaka University and RIKEN in Japan consists in reducing the number of physical qubits required to build logical qubits with using corrected and precise analog phase rotation gates involving a low overhead correction scheme instead of constructing it with costly combinations of error-corrected H and T gates[59]. This would enable the creation of useful early FTQC setups with only 10,000 physical qubits to support 64 logical qubits[60].

With ion traps, two challenges remain: keep improving qubit fidelities which are best-in-class and grow their qubit numbers.

The striking point is the sheer growth in qubit number that is needed to be able to create large-scale fault tolerant quantum computers able to break RSA 2048-bit keys using Peter Shor's famous algorithm and implement other (more useful) algorithms in the chemical simulation and optimization domains.

We need to increase the number of (high) quality qubits by 5 to 6 orders of magnitude. To implement a more reasonable 100 logical qubit system, we'd need an increase by only about 3 orders of magnitude.

You may think this shows a pessimistic view of the whereabouts of scalable quantum computers. The reality may be even darker. The fidelities from my charts use vendor data using very favorable benchmarking protocols, a bit like when a laptop vendor tells you their battery will last over 10 hours when in real life, it won't exceed 3 to 4 hours. Large discrepancies exist between qubit fidelities announced by vendors and those that are measured by users accessing the systems online on Amazon and Microsoft clouds and with IBM Quantum Experience online.

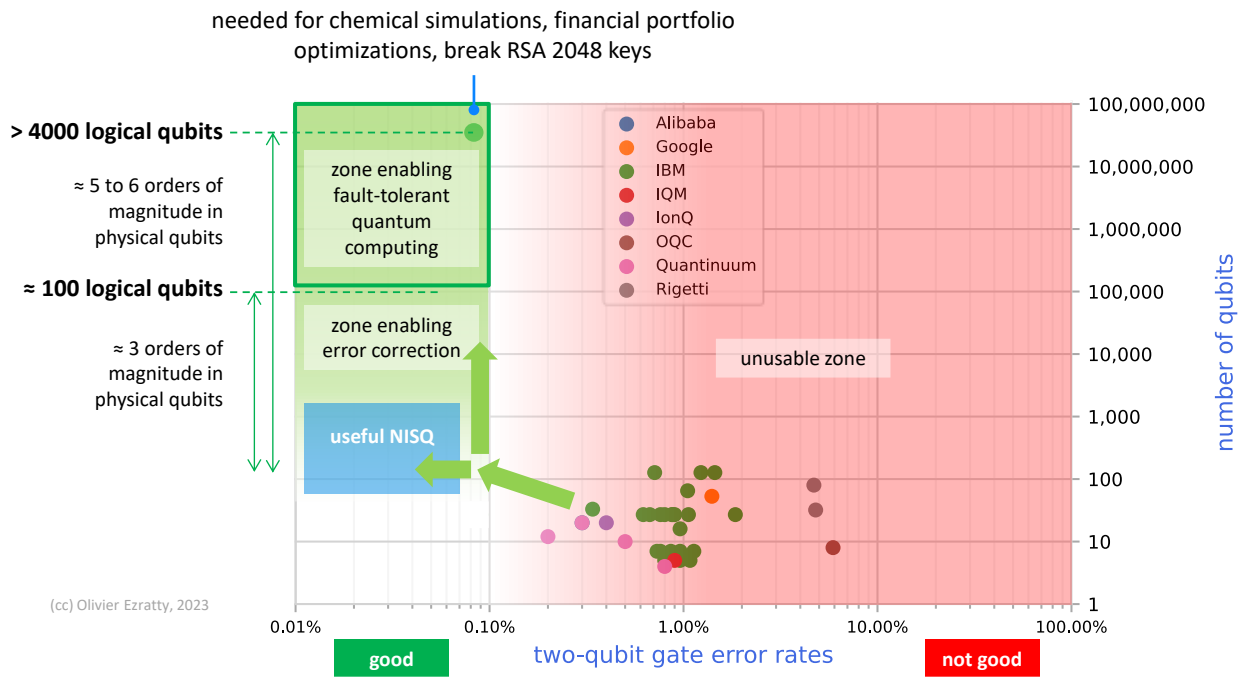

Figure 11: the paths to FTQC and NISQ. The number of physical qubits per logical qubits taken here in the left of the chart is based on typical asumptions on algorithms depth, which condition this ratio. Source: vendor fidelities numbers compiled by Olivier Ezratty.

Real-life benchmarks should for example measure two-qubit fidelities when several of these gates are simultaneously executed on a QPU and with repeated cycles.

This is usually implemented with randomized benchmarks with a set of random circuits followed by a rewind of these circuits taking advantage of reversibility. Qubits that are supposed to be back in the |0⟩ state are measured and here goes the errors. Interestingly, the usual required qubit fidelity is about 99.9% for both usable NISQ in the 50-100 qubit range, and for building scalable FTQC QPUs (quantum processing units). The 99.9% is an important Grail of the quantum computing industry. And as far as we know, 22 million of qubits with such fidelities are required to break an RSA 2048 bits key[61]. It could decrease to a more reasonable number of 350,000 qubits with cat-qubits[58].

But as shown in Figure 11, reaching some quantum advantage with NISQ QPUs is also very challenging. It requires even better qubits than FTQC, but at a moderate scale.

*Pending question:* will researchers and vendors improve qubit fidelities and reach 99.9% or even 99.99% at scale? This shows the need for some exponential growths to create logical qubits and a critical mass of logical qubits.

**Quantum volume and Q-score**

Quantum volume is a gate-based quantum performance metric proposed by IBM that provides an integer number to measure the quality and performance of its qubits[62]. The number gives an indication of the breadth (number of qubits) and depth (number of cycles of quantum gates) that a quantum computer can run with providing accurate results in two-thirds of the cases.

A quantum volume is a power of 2 and grows exponentially as you add one qubit, by design. Still, the best way to estimate what is this number is to use its power or $\log_2$(quantum volume) as shown in the table in Figure 12. So, every time we can use safely another qubit for computing a number of cycles equal to the considered qubit number, the quantum volume is doubling. And here we are with an exponential law ala Moore's law.

Since the creation of this metric in its current form in 2019, it has doubled about every 6 months. What is interesting to watch is the ratio between the related number of qubits and the quantum volume qubits. It's a high ratio with trapped ions QPUs and a much smaller one with superconducting qubits.

It shows the harsh reality also presented in Figure 9 and Figure 11: QPUs with a large number of qubits which are usually superconducting based have a rather low quantum volume but scale progressively while high-quantum volume systems are trapped ions based but don't scale as fast.

| Year | Brand | Version | Hw Qubits | Log2(QV) | % |
|---|---|---|---|---|---|
| 2017 | IBM | Tenerife | 5 | 2 | 40% |
| 2018 | IBM | Tokyo | 20 | 3 | 15% |
| 2019 | IBM | Johannesburg | 20 | 4 | 20% |
| 2020 | Honeywell | | 4 | 4 | 100% |
| 2020 | IBM | Raleigh | 28 | 5 | 18% |
| 2020 | IBM | Montreal | 27 | 6 | 22% |
| 2020 | Honeywell | H0 | 6 | 6 | 100% |
| 2021 | IBM | Montreal | 27 | 7 | 26% |
| 2020 | Honeywell | H1-1 | 10 | 7 | 70% |
| 2021 | Honeywell | H1-1 | 10 | 9 | 90% |
| 2021 | Honeywell | H1-1 | 10 | 10 | 100% |
| 2022 | IBM | Manhattan | 127 | 6 | 5% |
| 2020 | IonQ | Aria | 32 | 22 | 69% |
| 2022 | Quantinuum | H1-2 | 12 | 12 | 100% |
| 2022 | Quantinuum | H1-1 | 22 | 13 | 59% |
| 2023 | Quantinuum | H1-1 | 22 | 15 | 68% |

Figure 12: logarithms of various QPUs quantum volume over time and compared to the number of available physical qubits. Hw qubits corresponds to the QPU qubits (hardware). Vendors data compiled by Olivier Ezratty.

One caveat of quantum volume is it can't be used for QPUs in the quantum advantage regime since it requires being able to run some emulation of its randomized benchmark to compare its results and the QPU results. This means that the quantum volume exponential will stop at around $2^{54}$. Full stop. If it reaches this threshold anytime.

Figure 13 shows how these quantum volumes are growing over time, the best-in-class being trapped-ion qubits, mostly from Quantinuum.

Another way to benchmark quantum computers is to use a set of application related metrics that are independent from the qubit architecture, computing paradigm and can work at any scale. QED-C is supporting a series of such application oriented benchmarks proposed by researchers from Princeton, HQS, QCI, IonQ, D-Wave and Sandia Labs in the USA[63]. It mixes the volumetric benchmarking method from IBM and a comparison of performance with various standard algorithms.

Another architecture independent benchmark is the Q-score from Atos. It measures the maximum size of a standard optimization problem (Max-Cut) that can be solved on a given system[64]. As of early 2023, the best Q-scores were obtained on analog quantum computers (D-Wave[65], Pasqal although is a digital simulated way in that case[66]) in the 70-140 range. Gate-based quantum computers fare much lower, below 20. Still, we don't have enough history to learn how this Q-score evolves over time.

*Findings*: gate-based QPUs quantum volumes are currently quite low with no more than 23 qubits x 23 gate cycles. Application oriented benchmarks like Q-score tend to show that, currently, analog quantum computers fare better than gate-based QPUs.

*Pending question*: will this trend get reversed with gate-based systems faring better than analog quantum computers, and when?

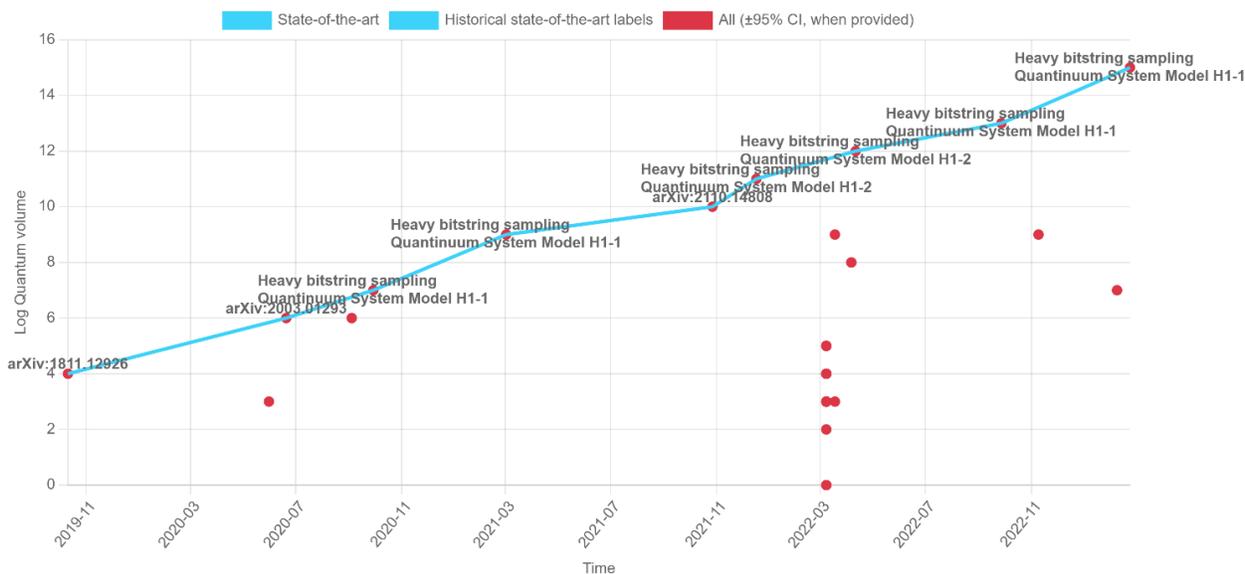

Figure 13: chart showing the evolution over time of the best quantum volumes around. Trapped-ions have been leading the pack for a while. Source: Unitary Fund Metriq initiative [57].

## EXPONENTIAL GROWTH POSSIBILITIES IN QUANTUM COMPUTING

We'll now try to classify the various technology challenges related to these various goals and those who could benefit from some real exponential progress and the others where it would be way more challenging.

Driving exponential growth with quantum computers to follow any equivalent of Moore's law is way more subtle than just improving chipset density like what was done during the first 50 years of the semiconductor chipsets industry. Quantum computers are not only diverse in nature but highly heterogeneous systems with many different parts that scale well or not with time. Quantum computing progress probably will not depend only on one critical component like the silicon chipset in classical computing, or similar technologies like memory and data storage.

Another debate is strong in the quantum computing community about the nature of the challenges ahead. For some, mostly physicists, we are still in the fundamental research stage and far from industrialization and the qubit quest is still a fundamental scientific endeavor. For others, we are already in the technology and engineering stages of this development. They are probably both right and wrong. We need science, technology, and engineering to create usable quantum computers. One cannot make it all alone.

Here, we will decompose the various scientific and technology challenges that researchers and industry vendors are facing and assess whether they could benefit from some exponential growth empirical laws or sometimes, specific breakthroughs instead of just some sustained continuous improvements.

### Qubit fidelities

As we have noticed in the charts above, improving qubit fidelities is the mother of all challenges in creating useful quantum computers. It is required both for creating useful NISQ quantum computers in the short/mid-term and to enable the development of FTQC (fault-tolerant quantum computers) in the longer term. The Holy Grail is to reach 99.9% fidelities for all operations: single and two qubit gates, and qubit readout, this last metric being important for fault-tolerant quantum computers using error correction codes.

The techniques used to improve qubit fidelities vary from one qubit type to the other. With so-called solid-state qubits (superconducting, quantum dots electron spin, NV or SiC cavities, topological), it depends a lot on the quality of the materials and manufacturing. There are also many interdependencies and compromises with qubit designs like the size, shape and arrangement of the bulky resonators and Purcell filters used to handle qubit states readout. As shown in Figure 14, the efforts to improve solid-state qubit quality also rely on improving the purity of the materials used (aluminum, niobium, tantalum, ...) and the quality of their deposition in the manufacturing process.

The choice of two-qubit gates can also impact their quality, and reduce the known crosstalk effect, when actions on some qubits disturb other distant qubits. For example, IBM is shifting CNOT gates to ECR (echoed cross-resonance) gates in their latest 127-qubit Sherbrooke processor. It reduces error rates, leakage and the need for frequent recalibrations.

With neutral atom qubits, fidelities depend on other factors such as the quality of the ultra-high vacuum in the atom chamber, which can be improved with using a 4K cryostat pulse tube cooling the vacuum pump, the quality and power of the lasers that control the atoms for their cooling, positioning and preparation.

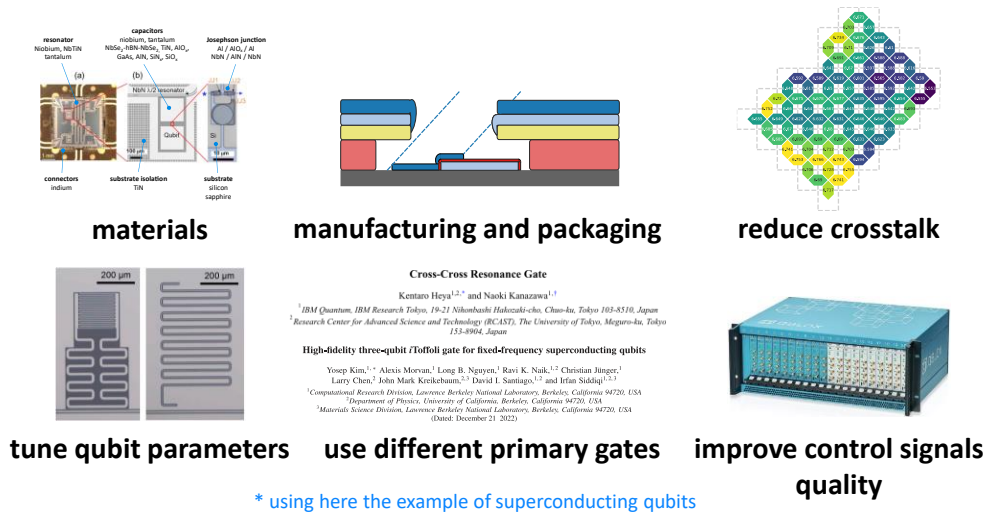

Figure 14: various means to improve qubit fidelities. Compilation: Olivier Ezratty.

In the photon qubit realm, challenges are different and are heavily related to the quality of individual photon sources and their ability to create large deterministic clusters of entangled photons. There are also challenges to scale the photon detectors used at the end of computing, when the photons have traveled through various optical devices (interferometers, polarizers, beam splitters, wave guides, etc.), usually integrated in nanophotonics circuits, mostly based on silicon.

Another way to ensure high qubit gates fidelities is to improve the quality of electronic drive signals as we'll see later, and in a scalable way.

Finally, in the solid-state qubit space, other types of qubits are designed to natively support in-hardware error correction schemes, called autonomous error correction. That is the case for bosonic and cat-qubits (Alice&Bob, QCI, Nord Quantique, Amazon) as well as with topological qubits (Microsoft). These better qubits are still in their infancy at this stage and have not yet been tested on a large scale, let alone on a... very low scale! But these qubits seem to be the only ones that could deliver one or two orders of magnitude of hardware-based fidelities improvements[67].

For the rest, improving fidelities while simultaneously growing qubit count seems to be an exponential problem. It relates to an overarching unanswered question: how far can we go to control a very large number of entangled quantum objects? Will quantum decoherence neuter that quest? Will we reach the quantum/classical frontier with such large quantum systems?

*Findings*: very unsure exponential growth, and it conditions all the rest!

*Pending questions*: what are the key scientific and technological fundamental bounds and drivers to improve qubit fidelities, qubit type by qubit type? Is it about better understanding the sources and effects of quantum decoherence or "just" an engineering problem?

**Qubit manufacturing and density**

Moore's law was above all enabled by the many advances in semiconductor manufacturing, mostly in UV/EUV photolithography and epitaxial deposition (PVD/CVD).

All sorts of manufacturing processes are at play with solid-state qubits.

**Superconducting qubits** chipsets are created in rather small fabs costing less than $50M and that are commonplace in many countries. You find them in research labs, Universities and vendors like IBM, Rigetti and Google have their own in-house fabs. They use electron beams lithography on single-chipsets wafers and sputtering matter deposition (not epitaxial). These relatively small fabs have a high turnaround and can produce a sample in about a week. The Josephson junction at the heart of these qubits consists mainly of an aluminum/aluminum oxide/aluminum (Al/AlOx/Al) sandwich created by double-angle evaporation of aluminum (aka sputtering) in vacuum with, in between, a controlled oxidation creating a tunneling barrier.

The complete creation cycle for a new chipset is about one month long, which is very fast in that industry. It enables fast test and error cycles. There were recent tests of superconducting qubits chipset manufacturing using 300 mm wafers and epitaxy, by IMEC, but it didn't generate better results than classical manufacturing methods using sputtering and non-epitaxial deposition[68].

**Quantum dots electron spin qubits** are manufactured in 300 mm wafer sized fabs with high quality standards and use epitaxial material deposition that generate high quality chipset features. This is done at a much higher cost than superconducting qubit manufacturing and a much slower turnaround. It can last over 1.5 years in total. This may explain why progress with these qubits is much slower than with superconducting qubits. Fabs doing this are with IMEC and CEA-Leti in Europe, Intel in Oregon (Hillsboro), Global Foundries in Malta, upstate New York State, IBM in Yorktown, etc. Surprisingly, the effect of Moore's law in this place may slow down qubit progress! Recently, both Intel and CEA-Leti acquired a cryoprober from Bluefors/afore which enables whole silicon wafer testing at 4K, helping shorten the characterization process. What is being said here is that, once qubit quality and control works, it will scale fast. Indeed, theoretically, Moore's law could be potentially implemented very fast, given the maturity of existing manufacturing capabilities, provided these qubits work well at scale.

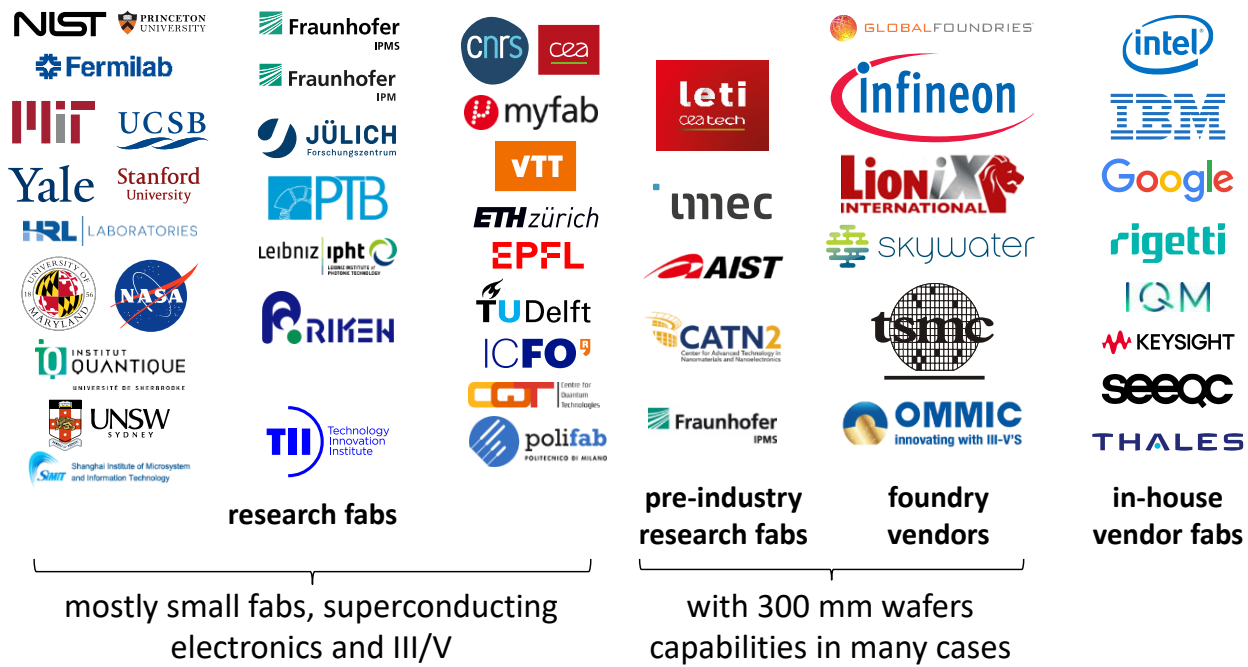

Figure 15: some quantum processor fabs in the world, from research to large scale industry production. This covers superconducting qubits, silicon qubits and some III/V photonics fabs. Compilation: Olivier Ezratty.

Another avenue in the spin qubit domain are carbon nanotubes from C12 Quantum Electronics[69] [70] (France) and carbon nanospheres from Archer Materials (Australia).

**Nitrogen and silicon carbide vacancies qubits** are less fashionable for quantum computing. They also face their own manufacturing challenges since it must be controlled at individual atom/vacancies levels. It seems difficult to create a large number of these defects. This is probably one of the few areas where a significant manufacturing challenge remains.

**Nanophotonics** chipsets are usually built on silicon in research or industry fabs. For example, PsiQuantum is partnering with GlobalFoundries for the manufacturing of its photonic qubit chipset, and Quandela is teaming with CEA-Leti in Grenoble. It seems that the scalability of photon qubits is more dependent on photon sources and detectors, which can potentially be incorporated in the main photonic chipset. These are usually produced in III-V technologies using compounds like gallium-arsenide in relatively small dedicated fabs. These are mastering the process of repetitive small layer deposition and ion milling process to structure and shape stacked quantum dots.

But with photon qubits, the real exponential progress that is waited for is the creation of large clusters of entangled states. In between, we can bet on multimode photons which can increase the computing Hilbert space at a reasonable hardware cost.

**EDAs**, the electronic design automation software tools running on classical computers are unsung heroes of semiconductor advances. They kept pace with the growth of chipsets complexity and come from Cadence, Synopsys and Siemens (formerly from Mentor Graphics) to name a few vendors, in a market that is being consolidated. I often hear the complaint that there are not enough of these tools available to design quantum chipsets, particularly with superconducting qubits and superconducting SFQ electronics[78]. One famous tool to design and even test your home-made superconducting chipset comes from IBM, Qiskit Metal (announced in 2021 and currently in alpha version[71]). In February 2023, Amazon AWS introduced Palace (PArallel, LArge-scale Computational Electromagnetics), a finite element open source code for full-wave electromagnetics simulations capable of simulating a single transmon qubit with its readout resonator coupling and a terminated coplanar waveguide (CPW) transmission line for input/output[72].

Cadence tools are used to design spin qubits and control electronics chipsets like with IQM[73]. Qubit designers usually assemble their own suite of multiscale EDA tools ranging from the physics of the qubits at low level to the whole behavior of a circuit. Tools frequently mentioned include Sonnet ABS (Adaptive Band Synthesis), Ansys HFSS 3D High Frequency Structure Simulation Software Multipurpose that helps the design and simulation of high-frequency electronics and Ansys RaptorQu which is used for electromagnetic simulations of superconducting qubit circuits. Drawpy is a Python framework based on HFSS that creates circuits drawings. At a higher level, Synopsis Technology Computer-Aided Design (TCAD) helps design a whole circuit. One challenge here is to create software suites able to modelize in a full-stack manner digital twins of qubit chipsets, from the lower quantum physics level up to all the circuitry and including noise models.

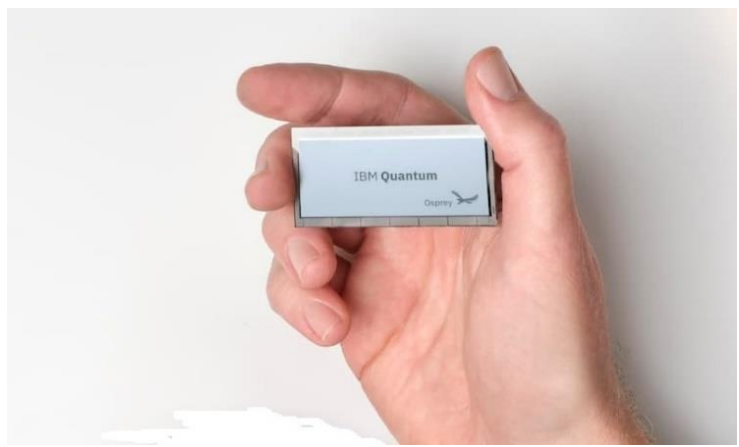

Figure 16: IBM's Osprey superconducting qubits processor shown in november 2022 and manufactured out of a 8 inch wafer. Source: IBM.

With regards to density, solid-state qubit size is not really constrained by manufacturing capabilities but more by the overall design and constraints of the qubits. For example, with superconducting qubits, the largest part of the qubit is not the Josephson junction where the quantum magic happens but with the surrounding resonators and capacitances. The resonator size is conditioned by the qubit drive microwave wavelength which is around 5 GHz for various reasons. Thus, the need for a quarter-wavelength 1.25 cm long resonator which consumes significant real estate on a chipset, even when squeezed in a packed serpentine.

The capacitors are bulky as well. Even with techniques like the coaxmon from OQC (UK) where the resonator is placed above qubits in a different chipset, the problem remains the same. A superconducting qubit is quite large, with at least a square of several millimeters wide. Some work is being done to miniaturize resonators and capacitances.

As a result of the size of qubits, IBM's Osprey 433 qubit chipset is already larger than most CMOS processors as shown in Figure 16. Thus, IBM's plan to assemble several of these processors and to interconnect them using short-range and longer-range microwave guides.

For silicon and photonic qubits, various figures of merit of the qubits and light processing also impact the size of the various parts of the chipset. And Moore's law has already been in play and can enable mass production of qubits in quantity, when it works. Silicon qubits are quite small, about 100 nm x 100 nm, and it would be relatively easy to scale their number on a given chipset to millions if not billions.

Another parallel can be made, this time with the "More than Moore" recent trend. More than Moore techniques are already implemented with qubit circuits designs with two (Google) or three (IBM, OQC) layers of chipsets packaging. These are separating qubit Josephson junctions and surrounding circuitry, the resonators for qubit readout and the rest of the wiring.

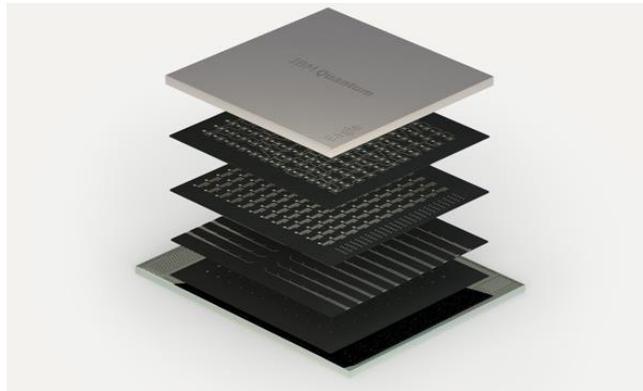

Figure 17: the three stacks in IBM's Eagle 127 qubit chipsets. Source: IBM.

Finally, manufacturers and industry vendors care about manufacturing quality and yield. The existing production volumes are still very low. The larger the fab, the smaller the defects are supposed to be. I have not found any data here.

*Findings*: chipsets manufacturing can have an impact on chipsets' quality. Density is more of a quantum physics problem than a manufacturing problem. There are still real limitations to manufacturing when materials must be deposited or removed at the individual atom level, as in the case of NV centers.

*Pending questions*: will qubit size and density be such a problem? Will it be possible to interconnect several QPUs with a small overhead?

**Gate and readout speed**

Although quantum computing is theoretically bringing some exponential speedup advantage compared to classical computing, its actual speed is important and can determine at which stage exactly, with a given algorithm and configuration, some speedup will be observed.

Qubit gates speeds are very different depending on qubit type. Superconducting gate speed ranges from 10 ns to about 400 ns depending on their type and microwave drive frequencies. It's constrained by being around 5 GHz for transmon qubits and by lower frequencies for fluxonium qubits, in the 1 GHz to 3 GHz range.

Since 2021, IBM measures the speed of its quantum processors with CLOPs, circuit layers per second, which nowadays ranges from 850 to 2.400 with plans to exceed 10.000.

IonQ trapped ions gate speed lasts between 135 µs (single qubit gate) and 600 µs (two-qubit gate)[74]. All in all, they are one thousand times slower than superconducting qubits gates. Of course, you must also evaluate the ratio between qubit lifetimes ($T_1$, $T_2$) and gate times, which is better for trapped ions.

For photon qubits, gate speed doesn't have much meaning since it's linked to the speed of light traversing various optical devices. Photonic qubits are fast by design.

Qubit readout speed is also important, particularly with non-demolition measurement (QND) that is being used with qubit types that enable quantum error correction. A non-demolition measurement collapses the wave function of the qubit on one of its basis states ($|0\rangle$ or $|1\rangle$ for a qubit, or more basis states for a qutrit or a qudits) whereas a demolition measurement "destroys" the qubit, like with a photon detector or a CCD based readout of a set of neutral atoms using fluorescence.

With superconducting qubits, readout takes several hundreds of ns. There, a compromise must be made between the readout speed which conditions the efficiency of error correction and the ability to multiplex several readout microwaves on a single cable. Using modern low-noise amplifiers at the 15 mK cold plate stage nearby the qubit chipset processor (TWPA for traveling-waves parametric amplifiers), several microwaves can be frequency multiplexed using a wide 2 GHz bandwidth with about 200 MHz per qubit. The shorter the microwave pulse, the broader will its spectrum be and vice-versa. You can't have it both ways. As a result, it is commonplace to multiplex not more than 10 qubit readouts on a single coaxial cable.

For comparison, a typical CMOS transistor commutation speed is about 3 ns. But of course, it only handles simple bits, not the wealth of analog and superposed information handled by qubit registers in QPUs.

*Findings*: gate and readout speeds are important figures of merit, in the NISQ regime for algorithms that are approaching a quantum-advantage and, later, for FTQC algorithms requiring a large number of logical gate operations.

*Pending questions*: what qubit technology has a potential to significantly improve its gate and readout operations?

**Qubit connectivity**

In solid-state qubits, their connectivity plays a key role for scaling quantum computing capacity. Even though D-Wave's chipsets have a best-in-class connectivity with their logical qubits containing physical qubits connected to 15 others, it is still not enough to ensure a generic quantum advantage with quantum annealing[75]. Gate-based superconducting qubits have at best a one-to-four connectivity (Google) as shown in Figure 18.

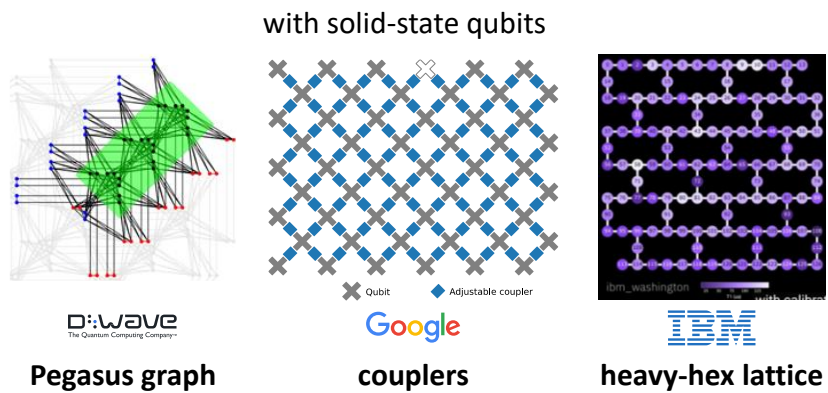

Figure 18: various types of qubit connectivity with superconducting qubits. Sources: D-Wave, Google and IBM.

IBM's qubits have a one-to-two and one-to-three connectivity in their "heavy hex lattice" layout. Improving this connectivity is not just a matter of stacking metal layers on top of the other like in classical CMOS chipsets. There are many constraints here with the wiring, undesired electromagnetic parasitic effects and the likes.

Still, IBM, the MIT Lincoln labs and others are working on multilayer connectivity chipsets with between 3 and 7 metal layers. This can for example enable long distance qubit connectivity which, in turn, can enable quantum error correction codes like the more efficient LDPC (Low-Density Parity Check) codes[76].

With ion traps, connectivity is supposed to be many-to-many but it doesn't seem to scale well to a large number of qubits (>40).

Improving qubit connectivity could have an enormous impact on the capacities of solid-state quantum computers. First, algorithms can run faster since there would be less need for SWAP gates when distant qubits have to be entangled together for whatever reason. Also, it could help create error correction codes and logical qubits requiring fewer physical qubits.

*Findings*: improving qubit connectivity could provide some levels of improvement in quantum computing capacity. But like with qubit fidelities, connectivity improvement is more a quantum problem than a simple electronics density problem.

*Pending questions*: what are the reasons, in terms of quantum physics and technology, why it is difficult to improve the connectivity of qubits, as with superconducting qubits?

## Qubit control

Qubit electronic controls play a key role in a quantum computer, sitting in between the driving classical computer and the qubits. In most installations, control electronics occupy a significant share of the quantum computer volume and power drain.

As shown in Figure 19, many control-related technologies play a key role in QPUs, and they have their own innovation cycle to drive scalability particularly with fault-tolerant quantum computer which require a very high number of physical qubits per "corrected" logical qubits.

**Electronics signals quality** have an impact on qubit gate fidelities. With microwave pulses, phase, frequency, and amplitude jitter can affect it, but no exponential progress is to be expected here. One related question is how electronics signals quality will not be degraded as vendors switch from high-quality high-power consumption lab-grade equipment to high-volume miniaturized equipment with potential challenges on signals quality.

**Electronics integration** of these electronic systems can be improved both at ambient temperature (FPGA could be replaced with ASIC) and low temperature (CryoCMOS, SFQ superconducting electronics, and spintronic)[77]. One key challenge here is to reduce power consumption. It can reach 100W per physical qubit and should be reduced to a couple mW. Many techniques are investigated that can enable this progress. One other key aspect is the capability to handle quantum error correction real-time decoding, under 1 μs for superconducting qubits[78]. We could expect some "Moore's law effect" here, that would certainly be related to the economics of this industry. One less known option is to implement cryo-control electronics with reversible logic[79]. One driver of electronics integration is a volume market. R&D poured in this will depend on a chicken-and-egg problem, driving the commoditization (or not) of quantum computers. Or the wealthy like IBM will continue to invest in-house efforts to make progress here.

**More than Moore** approaches are at play with heterogeneous electronic system designs combining at the "cold" level the qubit chipsets, qubit drive chipsets and, potentially, fiber optic/electronic conversion devices, and highly integrated circulators/amplifiers for qubit readout.

**Circulators and amplifiers** are indeed bulky passive and active devices that could be implemented in solid state circuits, favoring a much better integration. Many research labs are trying to miniaturize these devices with other designs. There are some paradigm shifts in the making here.

**Cabling and multiplexing** need to be miniaturized, otherwise scalable quantum computers will be very difficult to realize. Some acceleration of cables integration will need to show up. Then, multiplexing solution may be developed to circumvent the cabling and connectors mess in most solid state QPUs. IBM's recent Osprey "flexicable" was a significant progress for that respect, fitting the equivalent of 26 cables in a flexible that seemed to be only 10 cm large. Compared with the bulky cables previously fitting in cryostats, that was impressive.

**QND qubit measurement** (quantum non demolition). The way superconducting and silicon qubit are measured is incredibly complicated. A resonator received a shape microwave pulse that is reflected, traverse three circulators, then a first stage low-noise amplifier (at 15 mK for superconducting qubits), then another amplifier (at 4K), then an analog amplifier running at ambient temperature, then a mixer, then two analog to digital converters, then a FPGA or ASIC circuit using the ADC data (2 Giga-samples of 32 bits per seconds) analyze the resulting data to identify the readout microwave signal phase and amplitude to determine if the qubit is read as |0⟩ or |1⟩. A simpler solution would be welcome! But it's not just a matter of electronics. It's also a question of circuit quantum electrodynamics and various trade-offs between low-temperature and room-temperature electronics and cryogenics architecture.

**Neutral atoms** QPUs electronics are based on SLMs (Spatial light modulators, sorts of LCD chipsets controlling the phase of laser light used to position the atoms in space), AODs (acousto-optic deflectors), lasers and classical computing controls. Laser power and stability must grow as the number of qubits of these systems is increasing.

**Silicon nanophotonics** is another field of technological progress, mostly for photon qubits. These are related to a lot of research efforts to also embed photon sources and detectors in these chipsets, like what PsiQuantum is trying to do.

*Findings*: control electronics could benefit from some accelerated Moore's law driven by technology requirements and also by a shift of this market from an experimental one to a broad industry adoption. Like with Moore's law, progress will be driven by a combination of technology and economic incentives.

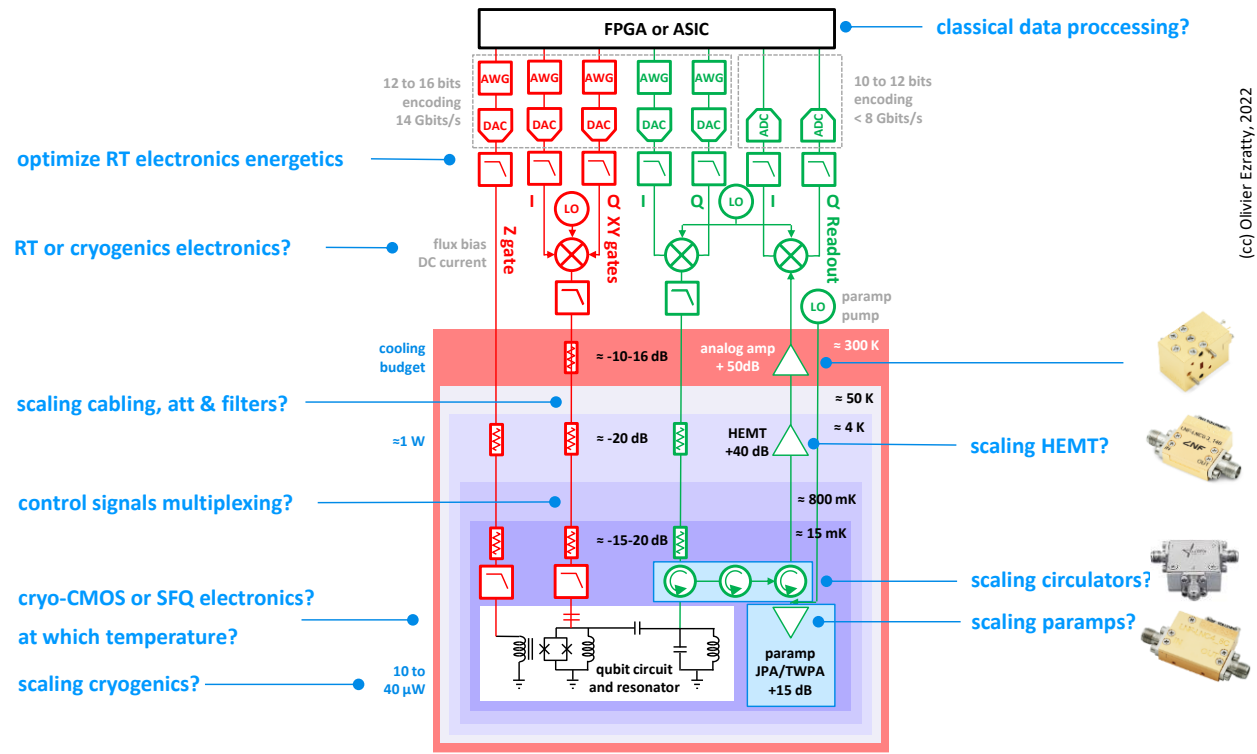

Figure 19: the various technology challenges in solidstate qubit enabling technologies. Source: Olivier Ezratty.

*Pending questions*: will it be possible, for solid state qubits, to fix either the cryoelectronics challenge or the signals multiplexing challenge? What are the theoretical lower bounds of qubit controls energetics? Could we suggest the creation of a "Quantum Dennard's scale" stating that control electronics and cryogeny could and should increase their qubit number support capacity with a stable energetic footprint?

## Cryogeny

Cryogeny is a rather classical technology, but it can benefit from dedicated R&D and show some sort of accelerated progress. So far, it has been obtained with brute force, packing more regular pulse tubes and dilutions in larger cryostats like with Bluefors's KIDE announced in 2021, and IBM's Goldeneye systems announced in 2020 and revealed in 2022[80]. Those systems are designed to host superconducting and silicon qubits up to a couple thousands, provided the cryo-electronics scales well at various levels (cabling, filters, attenuators, circulators, amplifiers).

Record breaking cryostats are also used in physics experiment outside the quantum computing world like with the one used with the CUORE 742 kg detector cooled at 10 mK that is looking for "neutrinoless double beta decay" (Figure 20).

But the largest cryostat being built so far is in the Department of Energy Fermilab in Chicago with a usable load of two meter in diameter and 1.5 meters height at 15 mK[81], shown in Figure 21. It is built to accommodate the very large prototype SRF superconducting qubits experimented by the Fermilab[82].

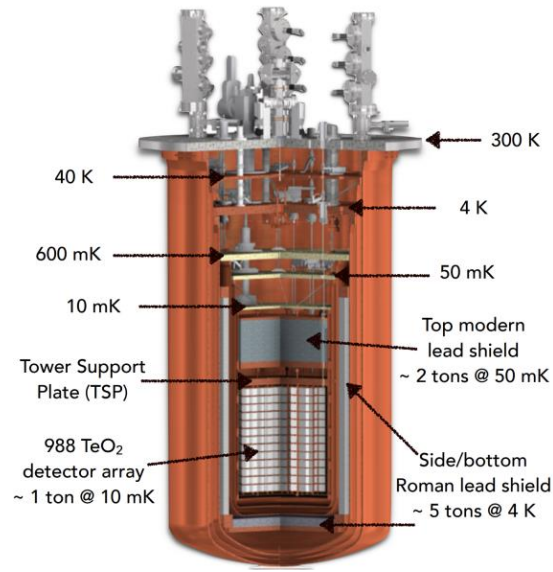

Figure 20: the CUORE mega-cryostat cooling a load of one ton. It is using dilutions from Leiden Cryogenics (The Netherlands) and compressors and pulse tubes from Cryomech (USA).

One key goal would be to improve the efficiency of these cooling systems to be closer to Carnot efficiency, both at the pulse tube (4K) and the dilution (15 mK) levels. It requires some innovations in thermodynamics engineering, one of them being the addition of more cooling stages[83]. It seems possible to significantly improve the efficiency of cryostats at 4K[84] and even below 1K[85]!

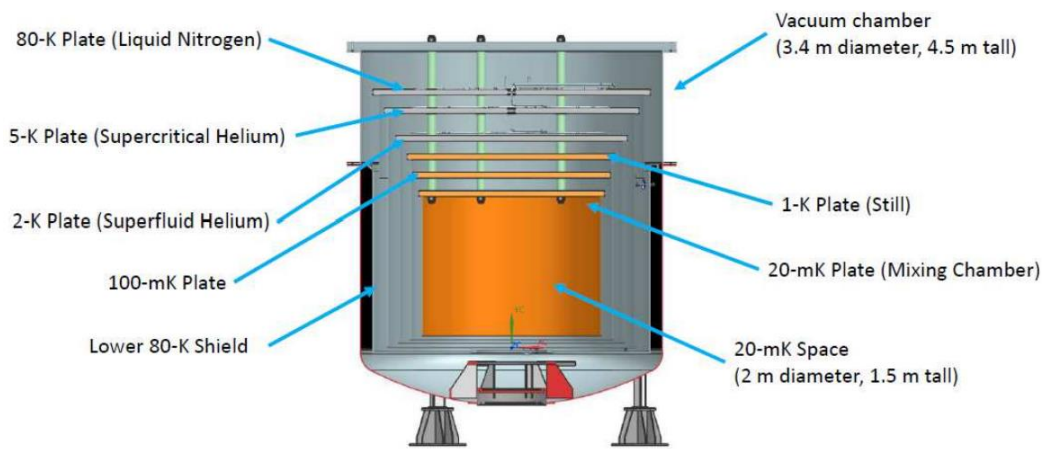

Figure 21: Fermilab's giant cryostat. Source: Fermilab.

But cryostats are used in most quantum computers. With photon qubit computers, they cool the photon sources and detectors at around 4K. As the number of photons increases, scaling with detectors will probably result in specific innovations at the related cryostat stage. They even cool ultra-vacuum pumps with neutral atoms QPUs.

*Findings*: cryogeny will have a significant impact on the ability to scale quantum computers in a reasonable way with fault-tolerant architectures both economically and on their energetics, these requiring the cooling of a large number of physical qubits and related electronics (cables, filters, attenuators).

*Pending questions*: will it be possible to get closer to Carnot efficiency in these cryostats?

### Other quantum computing paradigms

Until now, we've dealt with rather classical gate-based quantum computing systems, their figures of merits and how it could change over time. Other quantum computing paradigms abound which progress at different pace and under different constraints, with highlights on quantum annealers and analog quantum computers as shown in Figure 22. Hybrid quantum computing paradigms are also proposed like digital-analog quantum computing (DAQC)[86]. Also, measurement-based quantum computing (MBQC) is a variation of quantum computing adapted to flying qubits like photons. Although they can implement gate-based quantum computing models, topological qubits can also be defined as alternate quantum computing paradigms, at least at the low level.

We've already discussed D-Wave's quantum annealers and their steady 15-year long qubit number growing journey. Quantum annealers have some equivalents in the photonic space with coherent Ising machines[87], developed among others by QBoson, a startup from China[88], and NTT in Japan[89].

Analog quantum computers, also named programmable Hamiltonian simulators, usually based on neutral atoms and their high-energy Rydberg states, are closer than gate-based systems to reaching some form of quantum advantage. We have already seen that all these analog quantum computers have much better Q-scores than current noisy gate-based processors. We know about their vendors roadmaps.

D-Wave plans to reach 7000 qubits with better connectivity with its flux-qubit based quantum annealer while neutral atom vendors expect to reach thousands controllable qubits, even though you can't compare their number of qubits.

These systems have limits which are not well studied at this point. We still don't know how they will scale, how their "Quantumness" will work at large scale. And we know that these systems won't implement quantum error correction when run in analog mode.

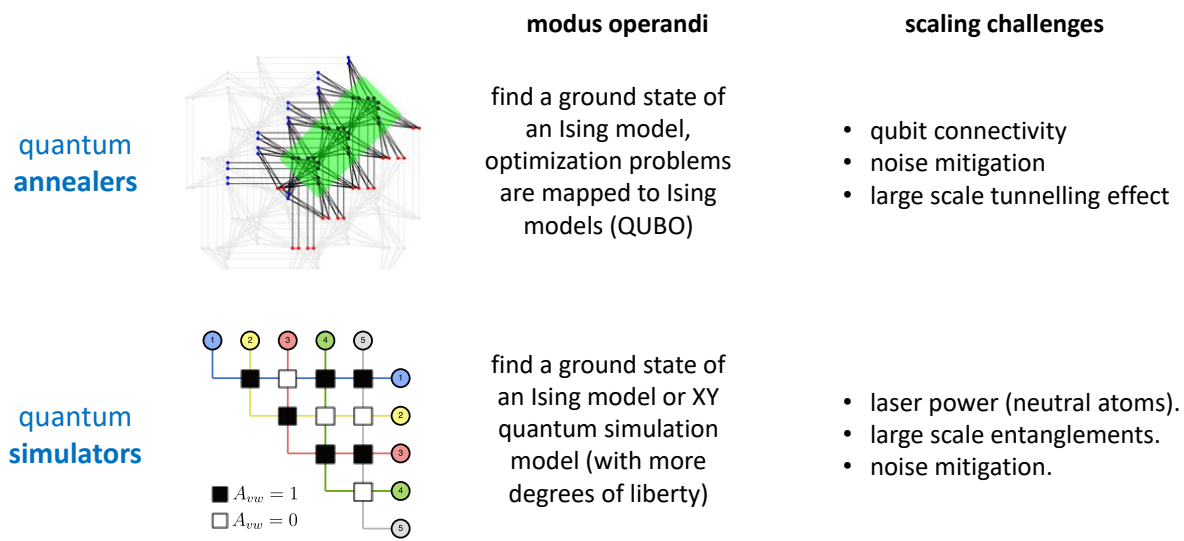

Figure 22: scaling challenges for some other types of quantum computers (quantum annealers, analog quantum computers). Source: Olivier Ezratty.

We then have topological qubits, mainly developed at Microsoft, with ups and downs, but they are still very serious about it. On an operational basis, they have the advantage of requiring simpler electronics for driving the qubits, using only DC current pulses, that are less costly to generate than microwave pulses. DARPA even selected this technology as well as neutral atoms from Atoms Computing and photons from PsiQuantum as part of their US2QC initiative which stands for Utility-Scale Quantum Computing and is looking at unconventional approach to create large scale quantum computers[90].

We could indeed put photonic-based systems in this "other paradigms" category, particularly MBQC systems, using large clusters of entangled photons and measurement-based computing paradigm. As stated twice already in this paper, their progress is mostly linked to the ability to create large ensembles of entangled photons cluster states. It is far too early to make forecasts on the progress made in creating these cluster states.

Photonic cluster states can be generated in many ways which have evolved over time with SPDC (spontaneous parametric down-conversion) using powerful laser single photons source heralding with a probabilistic outcome that is detected post-selectively and doesn't scale well beyond a dozen qubits[91], atom based cavity QED generation[92] that was later extended to ensemble of Rydberg atoms[93], individual neutral atoms, spin-photon entanglement to deterministically generate linear cluster states aka the Lindner-Rudolph protocol [94] with recent improvements[95], with quantum dots molecules[96], with the entanglement of several single photon sources[97], time-domain multiplexing using indistinguishable photon sources which has the advantage to be theoretically unlimited[98], 2D spin-photon cluster states[99], etc.

You can also add spectral domain multiplexing on top of time-domain multiplexing as a complement to SPDC sources[100]. The range of cluster sizes over 15 years of research spans 3 to 12 photon qubits. Here as well, scale is a challenge.

Quantum computing paradigms variations also include system architectures connecting quantum processing units and less demanding quantum memory units, significantly reducing the need for (logical) computing qubits[101].

<div align="center">**Systems architecture and software tools**</div>

Software is rarely mentioned when discussing Moore's law. However, most software sizes have grown over time as computing capacities have increased. Systems and application software also accumulate stuff over time since legacy code must often be preserved. Software size grows for several reasons: keeping legacy code, adding new features, added complexity and in some instances, improved developer productivity, enabled by more powerful development tools and also, simply, the growth of developer teams. And the fuss about "no code" is not changing that. The creation of higher abstraction level languages has not slowed down the trend.

I'll save you the many charts around on code complexity evolutions like those from fighter jets which indeed grew exponentially for several decades[102] or apples and oranges comparisons between operating systems, embedded software, more classical software and all software developed by Google[103].

Quantum computing changes the software equation a bit for many reasons. Here, I have more questions than answers. But they deserve to be asked!

- How will quantum code grow in complexity as quantum computers will be able to manage more qubits? It will be more complicated. We'll definitively leave the graphic circuit design realm. Python driven tools will dominate.

- What role system-level software will play in enabling or slowing down the qubit count growth?

- How will quantum software tools evolve? Will we have higher abstraction software development tools and what will be their (classical) computing overhead? Could they contribute to expanding the market reach of quantum computing?

- How will quantum error mitigation techniques and quantum error correction codes evolve? Particularly those who will relax the heavy constraints on hardware requirements and physical qubit fidelities[104].

- How will the connection between QPUs and driving classical computers improve to keep up with the tons of data to be exchanged? This is linked to the way quantum error correction will be implemented in FTQC systems. The goal is to have QEC being managed as close as possible to the qubits.

- How will developers assemble quantum software in serialized quantum algorithms and test it? Or will they just use spare typical quantum algorithms (HHL, QML variants)?

- How large quantum code will be debugged, certified and verified? This is an entirely new discipline being built at the research level.

- Will variational algorithms still dominate the field, even in the FTQC world?

- What will be the cost of the classical part of these algorithms, when they will build more complicated ansatzes or parametrized functions?

- How compilers, optimizers and transpilers will cope with the growing size of quantum code given quantum code optimization is a NP hard problem?

- Can we have sudden progress in quantum algorithmic design? It's possible even though a "Shor" moment is less probable.

- Can we have similar sudden progress in classical algorithmic design? In tensor networks computing? With dequantized quantum algorithms (quantum algorithms where an equivalent and efficient classical algorithm is found)? One recent example can be found with a recent work by Laura Lewis, a PhD from John Preskill, who created a classical machine learning algorithm to find the ground state of a many-body quantum system that requires only y $\mathcal{O}(\log(n))$ training data instead of $\mathcal{O}(n^c)$. That's a disruptive progress[105]!

*Findings*: software will play various roles in enabling many quantum computing technology developments.

*Pending questions*: see the above list.

## THE BUMPY ROAD FROM NISQ TO FTQC

There are many moments in the quantum computing roadmap ahead that may deliver some surprises, and also disappointments. Let's focus on the good surprises first.

One is that we may see some changes in the viability of various qubit types for both NISQ and FTQC QPUs. Cat-qubits, topological qubits and even photon qubits could surprise us. We will not shift from noisy quantum computers to magic fault-tolerant quantum computers doing everything spontaneously. This will be a progressive quest.

First, research labs and industry vendors must absolutely improve qubit fidelities and reach 99,9% at a decent scale, between 50 and 100 qubits. It is indispensable to build viable NISQ systems to run VQE (variational quantum eigensolvers, used to simulate quantum physics, the most promising avenue so far), QAOA (quantum approximate optimization algorithm, used for combinatorial optimization tasks) and QML (quantum machine learning) based algorithms which are all hybrid[106]. That's the most important challenge since, without it, nothing further down the road will be possible. If the early stage of NISQ with 50-100 qubits fails, FTQC will also fail. These are intertwined stories.

As we've seen before, then, the NISQ path can evolve with better qubits in the 100-1000s range while FTQC will require many more qubits but be satisfied with 99.9% fidelities. Which path is the easier route is not easy to guess. If and when physical qubits reach sufficient fidelities in the 99.5% to 99.9% range, then, we may see the first logical qubits show up with an improved fidelity vs their underlying physical qubits.

Then, two logical qubits and a corrected two-qubit gate will be implemented. Then, we'll have larger scale QPUs with tens of logical qubits, but with small improvement in fidelities, say, reaching 99.99% for example, enabling larger and deeper quantum algorithms. Further down the road, we'll see a synchronous growth of the logical qubit fidelities and their number. And, provided physical qubits can be safely assembled at larger and larger scales with keeping the base 99.9% fidelities.

We don't need to be faster with the logical qubit count vs their apparent fidelity since these figures of merit are interdependent.

Then we'll reach the 100 logical qubit threshold which will bring a real generic quantum advantage. Then, it will go beyond, with thousands of logical qubits and higher logical qubit fidelities, expanding the scope of fault-tolerant quantum computing.

The main condition is the ability of scientists to control large numbers of entangled qubits and maintain fidelities around 99.9%. It's often said that the theory says it's possible and that it is just an engineering problem. Some scientists disagree and think that ambient noise and complicated quantum interactions (mean fields, crosstalk, ...) will be showstoppers for this sort of scale. This may even turn out to be an exponential problem in itself!

The cross-over between NISQ and FTQC is due to the uncertainty between the emergence of scalable processors with over 99.9% fidelities qubits (FTQC) or small/mid-scale processors with higher fidelities (NISQ). You need higher logical qubit fidelities as you grow their number since the algorithm depth is correlated to the number of logical qubits, which in turn conditions the logical qubit target fidelity and the number of physical qubits per logical qubits.

As we've seen in the previous part, all this can happen through the accumulation of different technological development exponentials. In "*Optimizing resource efficiencies for scalable full-stack quantum computers*", Marco Fellous-Asiani et al proposed a holistic methodology able to combine the different technological stacks to optimize a quantum computer architecture and minimize its resource cost.

As an example, they figured out that a quantum computer factorizing RSA 2048 keys could be potentially built with a reasonable footprint and consume only about 40 kW[107]. This scenario was based on at least five advances of an exponential nature: qubit gate fidelities (>99.9% in a large and single monolithic chipset scale), surface code error correction with neglecting the energy required to decode the error syndromes, qubit control signals multiplexing (100 multiplexed qubit control drive microwaves in a single cable), cryogeny yield (close to Carnot efficiency), and control electronics efficiency (consuming about 1 mW per qubit, and using some assumptions on the microwave readout amplification chain). That's a significant alignment of planets which shows the enormous efforts at play in the realization of fault-tolerant quantum computers with a reasonable energetic footprint. A precise estimate of the energy required to decode the error syndromes, and the possibility to perform the decoding quickly enough will be a crucial aspect allowing to have precise estimate for the energy cost of surface code based quantum computers.

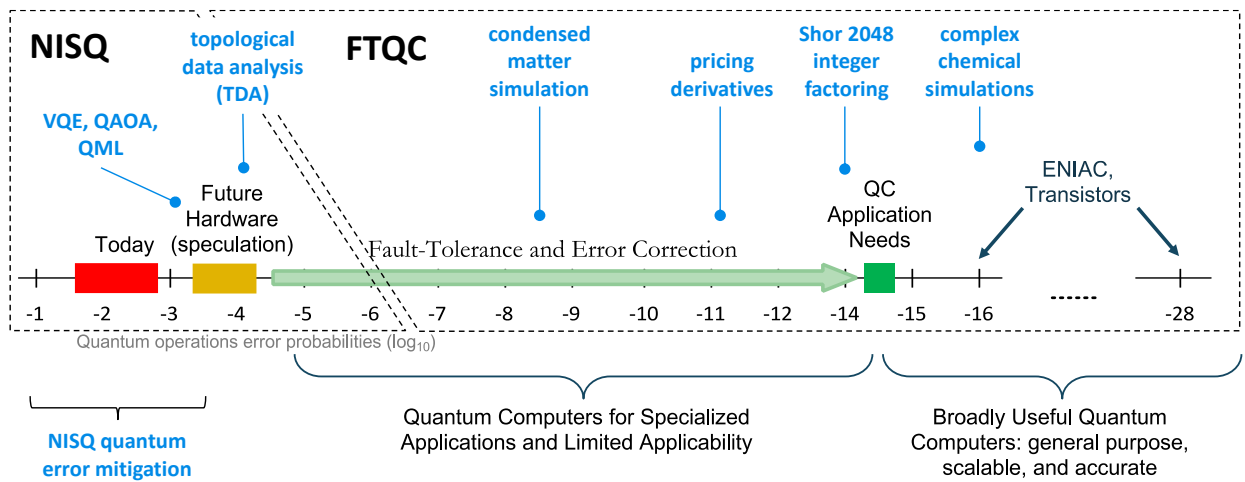

Figure 23: from NISQ to FTQC and logical qubit error rate requirements. Source: Bert de Jong, DoE[108], and additions by Olivier Ezratty.

**CHANGING FIGURES OF MERIT**

The initial business value of quantum computing was its potential capability to solve intractable problems with classical computing, in a reasonable human time. This potential seems still far away in decades. In the meantime, we'll maybe observe a shift in the business value brought by quantum computers. In the near term, they may bring some qualitative quantum advantage not related to some speedup and some energetic consumption advantage.

The energetics of quantum computing may also be favorable, first in the NISQ and analog regimes, and potentially, later in the FTQC regime. On NISQ, the idea was conveyed by John Preskill in his NISQ 2018 paper: "*Arguably, though, quantum technology might be preferred even if classical supercomputers run faster, if for example the quantum hardware has lower cost and **lower power consumption**"*[109].

Non-speed-up related quantum advantage can be many like a better quality of some QML classification and smaller training data sets requirements[110]. It may also come out with the variational quantum eigensolver algorithms used to compute chemical simulations in NISQ gate-based systems as well as their equivalents running on analog quantum computers[111].

That's what the Quantum Energy Initiative that I cofounded in August 2022 with Alexia Auffèves (CNRS MajuLab Singapore), Janine Splettstoesser (Chalmers University) and Robert Whitney (CNRS LPMMC Grenoble) is all about[112]. It ambitions to create a community gathering experts from various origins, from fundamental quantum physics to technology, from hardware to software, from research to industry, caring for the physical resource cost of emerging quantum technologies and willing to address the question in a scientific way.

This requires building new methodologies, language, roadmaps, metrics and benchmarking tools. And for both NISQ and FTQC era quantum computers, including using analog quantum computing paradigms (quantum annealers, analog quantum computers).

Another evolving set of figures of merit are the growing performance of classical supercomputers and their classical algorithms. Tensor network-based quantum inspired quantum algorithms are making huge progress. Even, the so-called traveling salesperson problem touted as an intractable problem for classical computers can be solved at very high scales with classical computers[113]. Classical heuristic-based algorithms and classical solvers or quantum inspired classical solvers like SQA (sequential quadratic approximation) are very efficient on problems with thousand variables on a simple PC. So, when looking at potential quantum speedups, one has to carefully look at best-in-class classical algorithms and hardware to make sound comparisons.

Finally, the quantum computing scene is getting fuzzier as some vendors are explaining that they are designing custom hardware tailored to solve specific customer problems. If implemented, these sorts of roadmaps will make benchmarking and accounting for performance even more difficult.

The more you deviate from a standardized computing model like what Moore's law and the microprocessor revolution enabled, the less likely you will create a strong market with a vibrant ecosystem. This would slow down any economic effect of an equivalent of Moore's law.

And for those who argue that it was the choice of the early mainframe computers in the 1950s and 1960s, let's remember a fact that they all died. IBM won the mainframe market with its generic platform family, the IBM 360, that covered both scientific and business computing needs. Its monopoly drove the Department of Justice to launch an antitrust suit in 1975 for a violation of the Sherman Act. It was withdrawn in 1982[114].

## DISCUSSION

Classical and quantum computing era advances are both driven by progress at the processor and chipset levels. In the case of quantum computing, we've seen the great variety of figures of merit and their interdependencies just at the qubit level. Quantum processors integration density is an important figure of merit, but not as much as qubit fidelities. This is a problem that the classical semiconductor industry was addressing smoothly. Nobody had to really care about error rates in classical CMOS processors even though it required some behind the scenes work. All in all, we need large numbers of high-fidelities qubits.

Their level of integration has however an indirect impact on the role of interconnect technologies, to create quantum links between quantum processors reaching their maximum size, which will be highly dependent of the qubit type.

Many different research organizations and industry vendors must fix seriously complicated problems. Is it more about science than engineering and technology development? All are necessary.

With regards to exponential laws ala Moore's law, quantum computing progress seems more chaotic. The best exponential progress may come from classical enabling technologies like control electronics, cabling and multiplexing and lasers. The zoology of qubit and qubit controls complexifies the landscape. We may also have some disruptions or "breakthroughs" coming from some qubit types, quantum error corrections and other clever system architectures. Then, we'll have to understand how both NISQ and FTQC scale up practically with potential conflicting scalability benefits, issues and overheads.

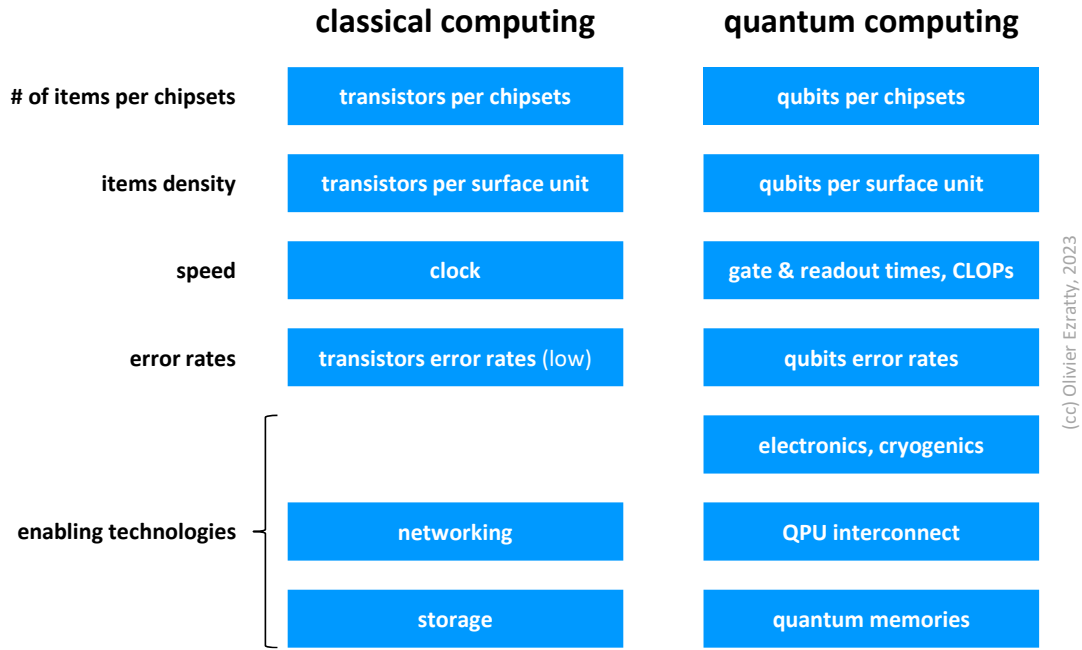

Figure 24: there are many similarities with the figures of merit of classical and quantum computers. Their progress is however more difficult in the case of quantum computing at the qubit quality level, and it depends a lot on the type of qubit. One key difference is the role of enabling technologies, particularly control electronics and cryogenics in the scaling of quantum computers. Source: Olivier Ezratty.

Also, charting exponentials ala Moore's law faces the lack of consistent end-to-end benchmarking. Comparing quantum computing and classical computing is particularly challenging as the latter is a moving target, and frequently associated with quantum computers when running hybrid algorithms. It is way more complicated than tracking the number of transistors per chipsets or their density.

The author warmly thanks Michel Kurek (Multiverse), Jean Senellart (Quandela), Marco Fellous-Asiani (Centre of New Technologies University of Warsaw), Reja Yehia (ICFO), Vincent Elfving (Pasqal) and Vivien Londe (Microsoft) for their feedback.

# Table of figures

30## Sources and notes